\documentclass[acmsmall,nonacm]{acmart}

\usepackage{multirow}
\usepackage{tabularx}
\usepackage{makecell}
\usepackage{placeins}
\setcopyright{none}
\newcommand{\ModelGrowth}{248}
\newcommand{\ComputeGrowth}{67}
\newcommand{\BandwidthGrowth}{47}
\newcommand{\CapacityGrowth}{42}
\newcommand{\InterconnectGrowth}{30}

\AddToHook{cmd/@mkabstract/after}{%
  \begingroup\small\smallskip
  \noindent GitHub: \href{https://github.com/Yufeng98/AI-datacenter}{\textcolor{blue}{\nolinkurl{github.com/Yufeng98/AI-datacenter}}}\par
  \endgroup}

\begin{document}

\title[Balancing Generality and Specialization in AI Datacenters]{Balancing Generality and Specialization: A Survey on AI Datacenter Hardware Architecture}
\author{Yufeng Gu}
\affiliation{%
  \institution{University of Michigan}
  \city{Ann Arbor}
  \state{Michigan}
  \country{USA}}
\email{yufenggu@umich.edu}

\author{Jiazhen Wang}
\affiliation{%
  \institution{University of Michigan}
  \city{Ann Arbor}
  \state{Michigan}
  \country{USA}}
\email{linusw@umich.edu}

\author{Reetuparna Das}
\affiliation{%
  \institution{University of Michigan}
  \city{Ann Arbor}
  \state{Michigan}
  \country{USA}}
\email{reetudas@umich.edu}


\begin{abstract}
Rapidly growing AI workloads are driving large investments in AI datacenters. This survey classifies industrial AI accelerators into four architectural categories and compares their compute and memory organizations. It examines how node-, rack-, and pod-scale interconnects support collective communication, and traces architectural evolution across accelerator generations. The analysis connects advances in arithmetic throughput with changes in precision, data delivery, execution coordination, communication, power delivery, and cooling. It also discusses future design challenges arising from workload diversity, data movement, infrastructure constraints, and model evolution, showing how the trade-off between generality and specialization extends from individual accelerators to datacenter-scale systems.
\end{abstract}


\maketitle

\section{Introduction}

\begin{figure}[ht]
    \centering
    \includegraphics[width=\linewidth]{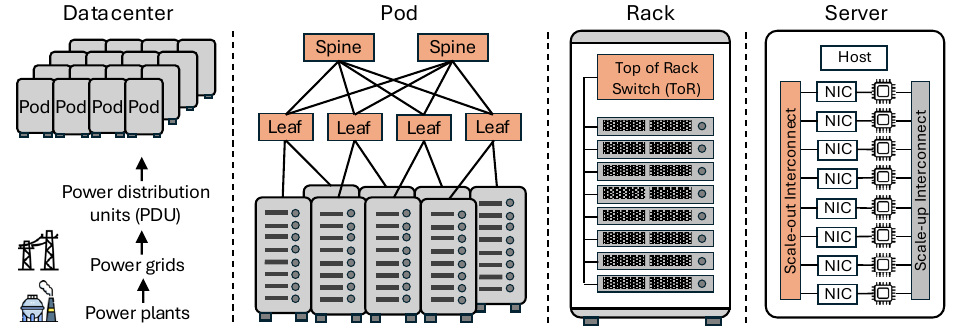}
    \caption{Architecture of datacenter, pod, rack and server.}
    \Description{Four panels zoom from a datacenter to a pod, rack, and server. Power plants feed the grid and power distribution units supplying the pods. Within a pod, spine switches connect leaf switches above racks. A rack contains servers attached to a top-of-rack switch. Within a server, accelerators connect to a scale-up fabric and to network interface cards on the scale-out fabric, with a host above them.}
    \label{fig:Datacenter_architecture}
\end{figure}

AI workloads execute on systems ranging from individual accelerators to datacenter-scale deployments. Large models may distribute both computation and model state across many devices, making execution dependent not only on the resources within each accelerator but also on the communication and system infrastructure that connects and supports them. Figure~\ref{fig:Datacenter_architecture} illustrates this physical hierarchy, spanning servers, racks, pods, and datacenters. At the server level, accelerators are integrated with local communication resources; racks and pods aggregate these servers into progressively larger systems, while datacenter infrastructure provides power delivery and thermal management. These physical levels, however, do not impose rigid communication boundaries, as scale-up fabrics may span multiple servers, racks, or even an entire pod. Large-model execution therefore engages every level of the hierarchy: computation and model state are distributed across devices, partitions exchange data over the interconnect fabrics, and facility-level power and cooling constrain how much of the installed hardware can operate concurrently.

Examining these levels together is increasingly important because the growth of AI model size can exceed the scaling of individual hardware resources. Across the selected observations in Figure~\ref{fig:scaling}, total model parameter counts grow at an endpoint annualized rate of approximately \ModelGrowth\% per year, compared with \ComputeGrowth\% for per-accelerator dense compute, \BandwidthGrowth\% and \CapacityGrowth\% for HBM bandwidth and capacity, and \InterconnectGrowth\% for scale-up interconnect bandwidth. Much of the growth in compute throughput is enabled by specialized arithmetic, including matrix engines and tensor cores, whereas memory and communication capabilities have scaled more slowly. Performance constraints therefore extend beyond arithmetic units to the broader hierarchy in Figure~\ref{fig:Datacenter_architecture}. This survey therefore examines industrial AI datacenter architectures across this hierarchy, with particular emphasis on how specialization and generality are balanced across accelerator architectures, interconnect technologies, and successive hardware generations.

Industrial AI accelerators address these increasing demands through several distinct organizations. Table~\ref{tab:accelerator_taxonomy} groups representative accelerator families into four categories. GPUs combine single-instruction, multiple-thread (SIMT) execution with dedicated tensor cores or matrix units. Neural processing units (NPUs) typically integrate matrix engines with vector, scalar, or other programmable execution units and explicitly managed local storage. Spatial dataflow architectures distribute computation and data across interconnected compute and memory resources, while compute-in-memory architectures place arithmetic units within or adjacent to memory arrays. These categories capture the dominant organization of execution and data movement rather than four uniform implementations.

\begin{table}[ht]
\centering
\small
\caption{High-level taxonomy of AI accelerator architectures.}

\label{tab:accelerator_taxonomy}
\begin{tabularx}{\linewidth}{%
    >{\raggedright\arraybackslash}p{0.14\linewidth}
    >{\raggedright\arraybackslash}X
    >{\raggedright\arraybackslash}p{0.32\linewidth}}
\toprule
\textbf{Category} &
\textbf{Computer Architecture} &
\textbf{Representative Platforms} \\
\midrule

\textbf{GPU} &
General-purpose SIMT architecture with dedicated tensor/matrix units &
NVIDIA GPU~\cite{nvidiaH100Paper}, AMD GPU~\cite{amdMI300XPaper} \\

\midrule

\textbf{NPU} &
Heterogeneous domain-specific architecture combining matrix, vector, and scalar engines around a shared on-chip scratchpad &
Google TPU~\cite{2023ISCATPUv4}, AWS Trainium/Inferentia~\cite{AWSTrainuim2,AWSInferentia2}, Qualcomm Cloud AI series~\cite{qualcommAI100}, Huawei Ascend~\cite{HuaweiAscend}, Intel Gaudi~\cite{IntelGaudi3HC36}, Microsoft Maia~\cite{MicrosoftMaia}, Cambricon MLU~\cite{cambriconBangCGuide} \\

\midrule

\textbf{Spatial Dataflow} &
Computation graph mapped onto distributed compute, memory, and communication resources; three styles: PE array, reconfigurable architecture, and functional-slice streaming processor &
\emph{PE array:} Tenstorrent~\cite{vasiljevic2024blackhole}, Meta MTIA~\cite{MTIA2}, Tesla Dojo~\cite{TeslaDojo}, Graphcore IPU~\cite{dissectingGraphcore}, Cerebras~\cite{cerebras2023}; \emph{Reconfigurable:} SambaNova~\cite{MICRO2024sambanova}; \emph{Functional-slice:} Groq~\cite{Groq2020TSP} \\

\midrule

\textbf{Compute-in-Memory} &
Memory-centric architecture that integrates compute logic in or near SRAM or DRAM modules to reduce data movement &
d-Matrix~\cite{dMatrixcorsair}, SK hynix AiM~\cite{AiMJSSCC}, Samsung PIM~\cite{Samsungaquabolt} \\

\bottomrule
\end{tabularx}
\end{table}

Designs within the same category can differ in supported operations, memory organization, and the balance between specialized and general-purpose resources. For example, GPUs retain general-purpose SIMT cores alongside dedicated matrix units, while NPUs such as Google TPUs and AWS NeuronCore combine specialized matrix engines with vector, scalar, or SIMD execution capabilities~\cite{nvidiaH100Paper,amdMI300XPaper,2023ISCATPUv4,NeuronCorev2}. Memory systems can exhibit a similar mixture, combining hardware-managed mechanisms with software-managed storage for explicit data reuse. Thus, the key distinction is not simply the accelerator label, but which computations and data movements receive dedicated hardware support and which remain under more general-purpose hardware or software control. Table~\ref{tab:accelerator_taxonomy} provides an entry point to the detailed comparison in Section~\ref{sec:taxonomy}, where these differences in compute engines and memory hierarchies are examined in greater detail.

At system scale, communication introduces architectural choices beyond the accelerator itself. Scale-up fabrics tightly couple groups of accelerators, whereas scale-out networks connect these groups across a broader cluster. The performance of distributed workloads depends not only on the physical network but also on how communication is scheduled over it. For example, an AllReduce operation specifies that participating devices combine their inputs and receive the reduced result, but it does not prescribe a particular ring- or tree-based schedule. Different schedules exercise the available paths differently, and their costs depend on factors such as message size, group size, and network topology~\cite{thakur2005mpich,nvidiaNCCLAlgorithms}. Relating collective algorithms to node-, rack-, and pod-scale fabrics therefore helps explain why the same operation may encounter different communication bottlenecks across system configurations.

Examining accelerator families across successive generations provides an additional perspective on architectural evolution. Changes in numerical formats or peak throughput capture only part of how one generation differs from the next. NVIDIA's Tensor Core evolution, for example, includes redesigned operand-delivery mechanisms and tighter coordination among execution units, enabling greater overlap between data movement and computation and supporting larger matrix operations~\cite{sun2022dissecting,luo2024benchmarking}. This generational view complements the architectural taxonomy by distinguishing features that persist within an accelerator family from those that change across releases. Accordingly, this survey traces the evolution of NVIDIA and AMD GPUs, Google TPUs, and AWS Neuron accelerators, including a detailed case study of NVIDIA's Tensor Core data path and control flow. The analysis considers not only advances in arithmetic capability but also accompanying changes in memory systems, communication infrastructure, and cooling requirements.

Together, these three perspectives motivate the challenges discussed in Section~\ref{sec:future}. Specialized hardware can execute target workloads efficiently but may be difficult to repurpose as demand shifts. Meanwhile, memory capacity, communication latency, power delivery, and cooling can constrain performance independently of workload evolution. These challenges highlight the tension between specialization and adaptability and the difficulty of supporting evolving algorithms within existing infrastructure.

This survey makes four main contributions:

\begin{enumerate}
\item We organize representative industrial accelerator families into a four-category taxonomy and compare their compute engines and memory organizations, providing a common framework for understanding how different designs execute operations and manage data.
\item We analyze node-, rack-, and pod-scale interconnect topologies together with collective communication algorithms, showing how communication schedules exploit physical connectivity and how their efficiency depends on workload characteristics and fabric organization.
\item We trace architectural evolution across accelerator generations, showing how increases in arithmetic throughput are accompanied by changes in precision, data delivery, execution coordination, communication, and deployment infrastructure.
\item We synthesize future design challenges arising from workload evolution, memory and communication constraints, power delivery and cooling, identifying implications for system flexibility and deployment longevity.
\end{enumerate}

Our investigation spans the broader industrial AI datacenter and accelerator landscape; the platforms analyzed in this survey are a representative selection from that study. We provide detailed comparisons of major platforms with sufficient public architectural information. The broader research corpus is publicly available at \href{https://github.com/Yufeng98/AI-datacenter}{\textcolor{blue}{\nolinkurl{https://github.com/Yufeng98/AI-datacenter}}}.

Section~\ref{sec:background} introduces the workload and infrastructure background and positions this survey relative to prior work. Sections~\ref{sec:taxonomy}--\ref{sec:evolution} present the accelerator taxonomy, scale-up interconnect analysis, and generational case studies, respectively. Section~\ref{sec:future} discusses future design challenges, and Section~\ref{sec:conclusion} concludes the survey.

\section{Background and Related Works}
\label{sec:background}

The scaling of modern AI workloads has shifted architectural design from individual processors toward the datacenter as a whole. Section~\ref{sec:industrial-scale} introduces energy demand, Section~\ref{sec:why-specialization} examines compute, memory, and communication scaling, and Section~\ref{sec:landscape} presents a timeline of accelerator disclosures. We then position this survey relative to prior work and state its scope and evidence conventions.

\subsection{The Rise of Industrial-Scale AI Infrastructure}
\label{sec:industrial-scale}

Recent advances in artificial intelligence have been driven not only by algorithmic innovation but also by an unprecedented expansion of compute infrastructure. Nearly 90\% of notable AI models in 2024 originated from industry, up from 60\% in 2023, and the training compute required by frontier models continues to double approximately every five months~\cite{stanford2025aiindex}. It is estimated that training compute for frontier language models has grown by approximately 5$\times$ per year since 2020~\cite{epoch2025trends}. As a result, progress in AI capability is increasingly coupled to the ability to provision and operate large-scale computing infrastructure rather than to improvements in single-processor performance alone.

This growing compute demand has made energy a first-order architectural constraint. The International Energy Agency estimates that global datacenter electricity consumption reached approximately 415~TWh in 2024 and could more than double to about 945~TWh by 2030, with AI representing a major source of growth~\cite{iea_energy_ai_exec}. In the United States, Lawrence Berkeley National Laboratory reports that datacenters accounted for approximately 4.4\% of total electricity consumption in 2023 and projects this share to rise to 6.7\%--12\% by 2028~\cite{shehabi2025lbnl}. These trends position AI datacenters not only as a computer-architecture challenge but also as an energy-infrastructure and deployment challenge, motivating the recurring emphasis in this survey on power delivery, cooling, and system-level resource constraints.

\subsection{Why Specialization Is Necessary}
\label{sec:why-specialization}

\begin{figure}[htbp]
  \centering
  \includegraphics[width=\linewidth]{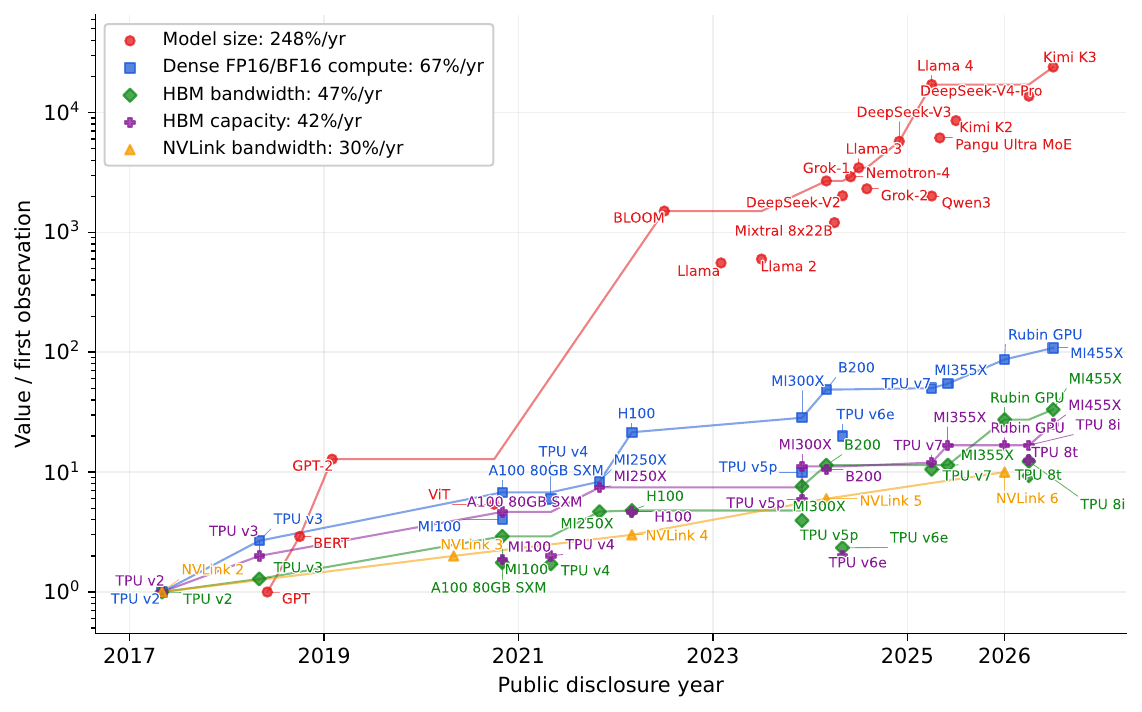}
  \caption{Normalized scaling of representative frontier-model parameter count, per-accelerator dense FP16/BF16 compute, HBM bandwidth and capacity, and per-GPU bidirectional NVLink bandwidth. Each series is normalized to its first observation; legend percentages are endpoint annualized growth rates between the first and last observation of that series. Model points use total parameter counts rather than active parameters.}
  \Description{A logarithmic plot against public disclosure date compares five independently normalized series from 2017 onward. The model-parameter series rises most steeply, followed by dense compute, HBM bandwidth, HBM capacity, and scale-up bandwidth. The displayed endpoint annualized rates are \ModelGrowth, \ComputeGrowth, \BandwidthGrowth, \CapacityGrowth, and \InterconnectGrowth{} percent, respectively. Every observation is labeled with its model, device, or NVLink-generation name in the corresponding series color; displaced labels connect to their points with thin leader lines. Individual observations vary around lines connecting record-setting values.}
  \label{fig:scaling}
\end{figure}

Figure~\ref{fig:scaling} compares selected model and hardware observations. Total model parameters have the largest endpoint annualized increase (\ModelGrowth\%/yr)~\cite{deepseekV4Pro,kimiK3}, followed by dense FP16/BF16 throughput (\ComputeGrowth\%/yr), HBM bandwidth (\BandwidthGrowth\%/yr), HBM capacity (\CapacityGrowth\%/yr), and the NVLink-only interconnect series (\InterconnectGrowth\%/yr).  Thus, model scale is increasing faster than compute capability, while compute capability is improving faster than memory and communication resources, placing increasing pressure on the broader system stack. Independent analyses of the memory wall report a consistent high-level trend~\cite{gholami2024memorywall}.

Accelerator architectures respond to these imbalances through coordinated advances in computation, memory, and communication. Specialized matrix engines and low-precision arithmetic improve the efficiency and throughput of common AI operations, while higher memory bandwidth and greater on-chip data reuse reduce the need to access external memory for every operand. High-bandwidth interconnects further enable model state and computation to be distributed across multiple accelerators, extending effective compute and memory resources beyond a single device. Because these mechanisms target different system bottlenecks, their relative emphasis varies across architectures, giving rise to the diverse accelerator designs introduced below and examined in Section~\ref{sec:taxonomy}.



\subsection{A Decade of Datacenter AI Accelerators}
\label{sec:landscape}

\begin{figure}[tbp]
  \centering
  \includegraphics[width=0.8\linewidth]{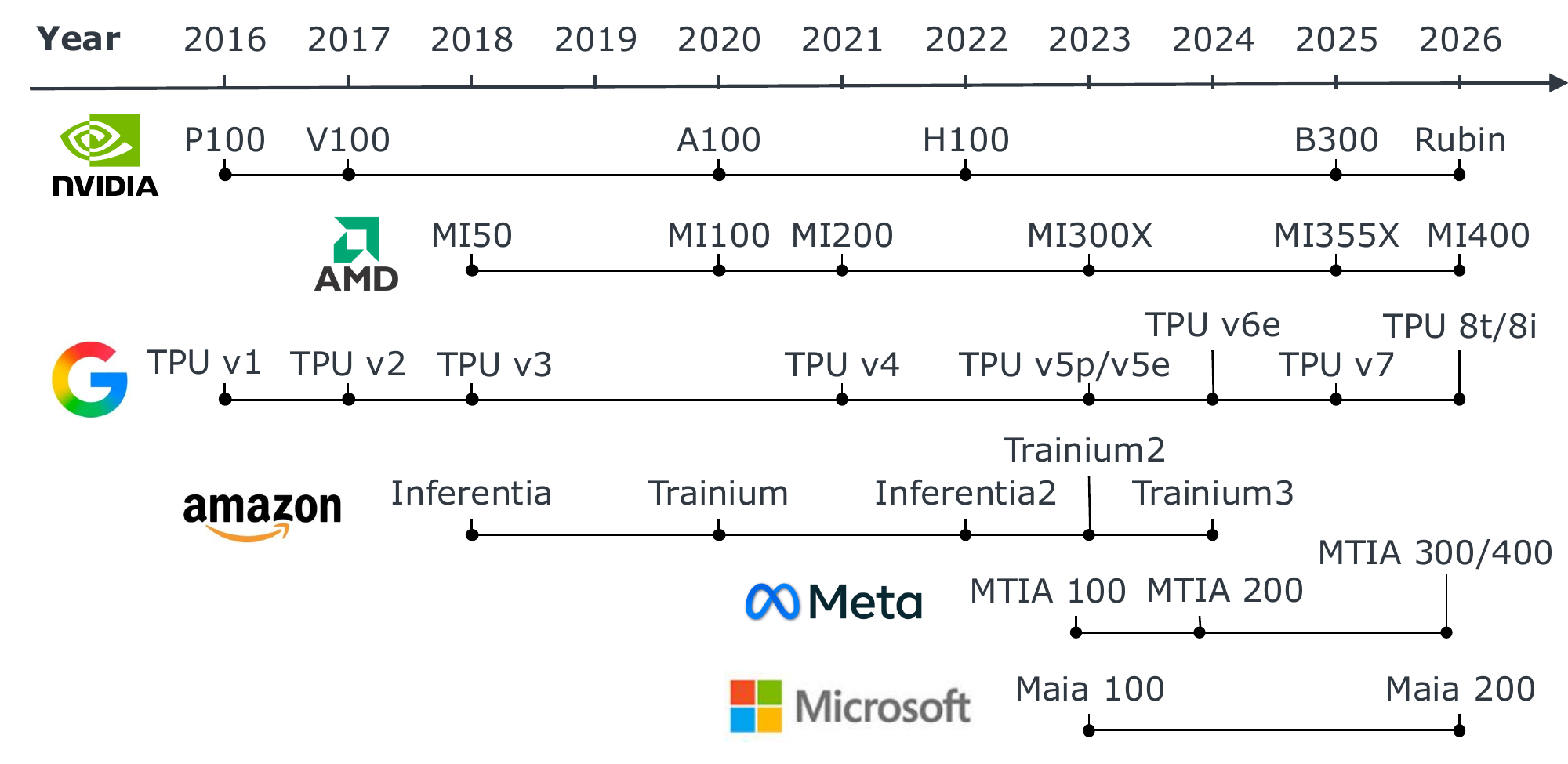}
  \caption{Selected AI accelerators disclosed from 2016 to 2026.}
  \Description{A horizontal year axis from 2016 to 2026 appears above six rows identified by company logos. From top to bottom, the rows show NVIDIA P100 through Rubin; AMD MI50 through MI400; Google TPU v1 through TPU 8t and 8i; Amazon Inferentia and Trainium on one line; Meta MTIA 100 through MTIA 300 and 400; and Microsoft Maia 100 and 200. Black circles mark public disclosure years. TPU v5p and v5e, TPU 8t and 8i, and MTIA 300 and 400 each share a label.}
  \label{fig:accelerator_timeline}
\end{figure}

Figure~\ref{fig:accelerator_timeline} traces selected accelerator generations from NVIDIA, AMD, Google, Amazon, Meta, and Microsoft. In parallel with continued GPU development, cloud and platform providers have introduced custom accelerators for the AI services they operate. Google publicly disclosed its TPU in 2016~\cite{googleTPUIntroduction2016}, AWS announced Inferentia in 2018~\cite{awsInferentiaIntroduction2018}, and Meta and Microsoft introduced MTIA and Maia, respectively, in 2023~\cite{MTIA,microsoftMaiaIntroduction2023}.

These accelerators reflect different optimization objectives. NVIDIA and AMD datacenter GPUs provide programmable platforms for a broad range of AI training and inference workloads, combining specialized matrix computation with general-purpose parallel execution~\cite{A100WhitePaper,amdMI300XPaper}. By contrast, in-house accelerators from cloud and platform providers are often tailored to the models, service requirements, and cost constraints of their own deployments, including the distinct demands of training and inference. AWS introduced Inferentia to reduce inference cost while maintaining high throughput and low latency, followed by Trainium for model training~\cite{awsInferentiaIntroduction2018,awsTrainiumIntroduction2020}. Google's first TPU targeted inference, whereas TPU v2 and v3 extended the family to training~\cite{TPUv2v3}. More recently, TPU 8t has emphasized large-scale pre-training, while TPU 8i has targeted post-training and latency-sensitive serving~\cite{tpuv8}.

The workloads shaping these designs also evolve across generations. Meta's MTIA 100 and 200 focused on ranking and recommendation inference~\cite{MTIA,MTIA2}; MTIA 300 extends this focus to recommendation training, while MTIA 400 broadens support to generative AI while retaining recommendation workloads~\cite{metaMTIAEvolution2026}. Microsoft's Maia family followed a different trajectory: Maia 100 was introduced for cloud AI training and inference, including OpenAI models and Copilot, whereas Maia 200 more explicitly targets the cost of token generation during inference~\cite{microsoftMaiaIntroduction2023,microsoftMaia200Introduction2026}. Section~\ref{sec:taxonomy} compares the architectural approaches used to serve these different objectives, while Section~\ref{sec:evolution} examines how those designs change across successive generations.

\subsection{Related Work}
\label{sec:related-work}

Several recent surveys cover overlapping ground. The Lincoln AI Computing Survey (LAICS) catalogs peak-performance and power-envelope data across edge and datacenter AI/ML accelerators~\cite{reuther2025laics}. Duan~et~al.\ survey efficient training infrastructure for large language models, focusing on parallelism strategies, scheduling, and fault tolerance for 10k-GPU-class clusters~\cite{duan2024training}. Li~et~al.\ examine LLM inference acceleration from a hardware perspective spanning CPU, GPU, FPGA, ASIC, and PIM/NDP~\cite{li2024llminfer}. Xu~et~al.\ provide the closest taxonomy-style survey of neural-network hardware, spanning GPUs, TPUs/NPUs, FPGAs, ASICs, in-/near-memory, and neuromorphic designs at both academic and industrial scale~\cite{xu2025nnsurvey}. Koilia and Kachris perform a normalized cross-platform comparison of LLM accelerators on a common 16~nm node~\cite{koilia2024llmhw}. Silvano~et~al.\ survey deep-learning accelerators for heterogeneous high-performance computing platforms in ACM Computing Surveys~\cite{silvano2025dlhpc}. Adjacent surveys address inference-engine software stacks~\cite{park2025engines}, hardware/software co-design across the LLM lifecycle~\cite{guo2024codesign}, and industrial collective-communication libraries~\cite{weingram2023xccl}. On the facility side, Chen~et~al.\ review the electricity demand and grid impacts of AI datacenters~\cite{chen2025griddemand}, and Bashir~et~al.\ argue for co-designing AI datacenters with the power grid~\cite{bashir2026gridcodesign}.

Our survey differs from these prior works in its combined treatment of five aspects: (1)~it introduces a taxonomy of industrial AI accelerators from the perspective of balancing specialization and generality (Section~\ref{sec:taxonomy}); (2)~it focuses on AI datacenter architecture, connecting node-, rack-, and pod-scale scale-up network topologies with the collective-communication operations mapped onto them (Section~\ref{sec:scaleup}); (3)~it traces the evolution of compute precision, execution models, memory hierarchy, and interconnect across accelerator generations (Section~\ref{sec:evolution}); (4)~it examines power supply and cooling as architectural constraints linking accelerator, rack, and facility design (Sections~\ref{sec:evolution_power_supply} and~\ref{sec:future_cooling}); and (5)~it discusses the challenges that evolving AI workloads and models pose for hardware specialization and future AI datacenter design (Section~\ref{sec:future}).

\section{Architectural Categories of AI Accelerators}

\label{sec:taxonomy}

\begin{figure}[h]
    \centering
    \includegraphics[width=0.98\linewidth]{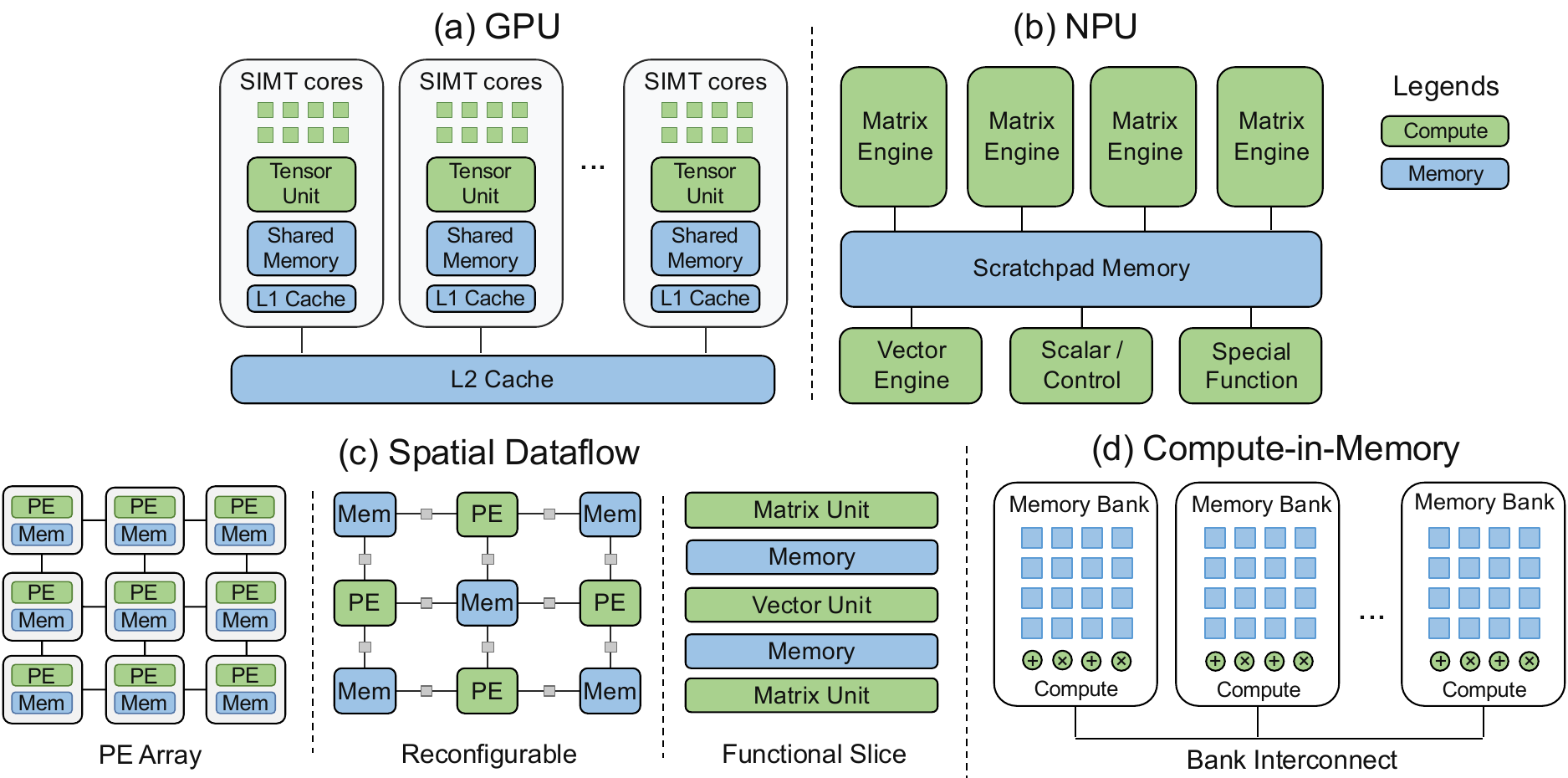}
    \caption{Architectural categories of AI accelerators. (a) \textbf{GPU}: SIMT cores integrated with tensor units and supported by a hybrid memory hierarchy that combines shared memory and caches. (b) \textbf{NPU}: Heterogeneous compute engines, including matrix, vector, scalar, and special-function units, organized around a shared scratchpad memory. (c) \textbf{Spatial dataflow} accelerators: Three representative substyles are illustrated, including PE array, reconfigurable fabrics, and functional-slice streaming processors. (d) \textbf{Compute-in-memory}: Computational operations are performed within memory arrays or in logic tightly coupled to memory banks.}
    \label{fig:accelerator_arch}
    \Description{Four architectural schematics use green for compute and blue for memory. GPU blocks combine SIMT cores, tensor units, local shared memory and L1 caches above a common L2 cache. The NPU places matrix, vector, scalar, and special-function engines around scratchpad memory. Spatial dataflow examples show paired processing and memory tiles, a reconfigurable mesh, and alternating compute and memory slices. Compute-in-memory banks place arithmetic beside memory cells and connect through a bank interconnect.}
\end{figure}

Specialized accelerators can provide substantially higher throughput and energy efficiency than general-purpose CPUs on AI workloads. Table~\ref{tab:accelerator_characteristics} summarizes the key characteristics of representative AI accelerators, and Figure~\ref{fig:accelerator_arch} illustrates their canonical architectural organizations. Broadly, these architectures can be classified into four categories according to the dominant architectural and programming idiom: (1) GPU-based designs that combine SIMT cores with tensor or matrix units; (2) heterogeneous NPU architectures that integrate matrix, vector, and scalar engines around a shared on-chip scratchpad memory; (3) spatial dataflow architectures that map a computation graph onto distributed compute, memory, and communication resources, spanning PE array, reconfigurable architecture, and functional-slice streaming processors; and (4) compute-in-memory architectures that embed arithmetic within or beside memory arrays. 

These categories represent different approaches to exploiting the inherent properties of AI workloads, including increasing data reuse, specializing compute resources, and improving data locality. In this section, we first introduce these four categories of AI accelerator architectures and then examine their compute engines and memory hierarchies in detail; Section~\ref{sec:scaleup} examines their scale-up interconnects. We assign each platform a primary category according to its dominant execution and data-movement model, while recognizing that secondary mechanisms such as systolic execution, compiler-managed placement, and near- or in-memory arithmetic can overlap across categories.

\begin{table*}[htbp]
\centering
\footnotesize
\renewcommand{\arraystretch}{0.92}
\setlength{\tabcolsep}{4pt}
\caption{Key architectural features of representative AI accelerators.}

\label{tab:accelerator_characteristics}
\begin{tabularx}{\textwidth}{@{}%
    >{\raggedright\arraybackslash}p{0.19\textwidth}
    >{\raggedright\arraybackslash}p{0.34\textwidth}
    >{\raggedright\arraybackslash}p{0.15\textwidth}
    >{\raggedright\arraybackslash}X@{}}
\toprule
\textbf{Accelerator} &
\textbf{Architecture} &
\textbf{Memory} &
\textbf{Scale-up Topology} \\
\midrule

\multicolumn{4}{@{}l}{\rule{0pt}{2ex}\textbf{Category 1: GPU --- SIMT and tensor/matrix cores}} \\
\midrule

NVIDIA GPU~\cite{nvidiaH100Paper,nvidiaRubin2026} &
SIMT + tensor cores &
HBM &
Cube mesh / switched all-to-all\textsuperscript{1} \\
\cmidrule(lr){1-4}

AMD GPU~\cite{amdMI300XPaper,amdMI455X2026,amdCDNA5} &
SIMT + matrix cores &
HBM &
Direct full mesh / switched all-to-all \\

\midrule

\multicolumn{4}{@{}l}{\rule{0pt}{2ex}\textbf{Category 2: NPU --- heterogeneous matrix, vector, and scalar engines}} \\
\midrule

Google TPU~\cite{tpuArch, 2023ISCATPUv4, tpuv5p, tpuv5e, tpuv6e, tpuv7, tpuv8} &
Matrix + vector + scalar; SparseCore or CAE\textsuperscript{2} &
HBM &
2D / 3D torus; Boardfly\textsuperscript{2} \\
\cmidrule(lr){1-4}

AWS Trainium~\cite{AWSTrainuim, AWSTrainuim2, AWSTrainuim3} &
Matrix + vector + scalar + programmable SIMD &
HBM &
2D torus; switched all-to-all\textsuperscript{3} \\
\cmidrule(lr){1-4}

Qualcomm AI 100~\cite{qualcommAI100} &
Matrix + VLIW vector and scalar units &
LPDDR &
\textemdash{} \\
\cmidrule(lr){1-4}

Huawei Ascend~\cite{HuaweiAscend, liao2025ub} &
Matrix + vector + scalar units &
HBM &
UB-Mesh: nD full mesh\textsuperscript{4} \\
\cmidrule(lr){1-4}

Intel Gaudi~\cite{IntelGaudi3HC36} &
Matrix + VLIW vector units &
HBM &
Direct full mesh \\
\cmidrule(lr){1-4}

Microsoft Maia~\cite{MicrosoftMaia,dighe2026maia200} &
Matrix + vector units &
HBM &
Switched Ethernet \\
\cmidrule(lr){1-4}

Cambricon MLU~\cite{cambriconBangCGuide,cambriconMLU270} &
Matrix + vector + scalar units &
DDR4 / LPDDR / HBM &
\textemdash{} \\

\midrule

\multicolumn{4}{@{}l}{\rule{0pt}{2ex}\textbf{Category 3: Spatial dataflow --- computation graph mapped onto distributed hardware resources}} \\
\midrule

Tenstorrent~\cite{vasiljevic2024blackhole,tenstorrentGalaxyMesh} &
RISC-V cores + matrix/vector units &
GDDR6 & 2D mesh Ethernet \\
\cmidrule(lr){1-4}

Meta MTIA~\cite{MTIA2,metaMTIA300ISCA2026,metaMTIAEvolution2026} &
PE grid with matrix + vector units &
LPDDR / HBM &
Switched PCIe / Ethernet-based fabrics\textsuperscript{5} \\
\cmidrule(lr){1-4}

Graphcore~\cite{dissectingGraphcore,graphcorePod64} &
Tile processors with FP16 vector MAC + FP32 scalar FPU &
DDR4 &
2D Torus \\
\cmidrule(lr){1-4}

Tesla Dojo~\cite{TeslaDojo} & Tile processors matrix + SIMD vector +
scalar units &
On-chip SRAM + HBM &
2D mesh die-to-die \\
\cmidrule(lr){1-4}

Cerebras~\cite{cerebras2023} &
Wafer-scale PE array; SIMD + scalar units; data-triggered execution &
On-chip SRAM &
2D mesh on-wafer\textsuperscript{6} \\
\cmidrule(lr){1-4}

SambaNova~\cite{MICRO2024sambanova} &
Reconfigurable SIMD compute + memory tiles &
HBM + DDR &
Direct peer-to-peer RDU links (SN40L) \\
\cmidrule(lr){1-4}

Groq~\cite{Groq2020TSP, Groq2022software} &
Matrix/vector units + SRAM/switch slices; statically scheduled &
On-chip SRAM &
Dragonfly \\

\midrule

\multicolumn{4}{@{}l}{\rule{0pt}{2ex}\textbf{Category 4: Compute-in-memory --- PIM and in-SRAM compute}} \\
\midrule

d-Matrix~\cite{dMatrixcorsair} &
Digital in-SRAM MAC array &
LPDDR5 &
Switched PCIe \\
\cmidrule(lr){1-4}

SK hynix AiM~\cite{AiMJSSCC} &
BF16 MAC-based PIM units &
GDDR6-AiM &
\textemdash{} \\
\cmidrule(lr){1-4}

Samsung PIM~\cite{Samsungaquabolt} &
FP16 SIMD PIM processors &
HBM-PIM &
\textemdash{} \\

\bottomrule
\end{tabularx}

\vspace{2pt}
\begin{minipage}{\textwidth}
\scriptsize
\textsuperscript{1} DGX-1 uses hybrid cube mesh; DGX-2 uses NVSwitch-based all-to-all~\cite{nvidiaP100Paper,nvidiaHGX2Topology,nvswitch}.\par
\textsuperscript{2} TPU engines vary by generation; 8i replaces SparseCores with a Collectives Acceleration Engine (CAE). Torus applies to supported slices; 8i uses Boardfly.\par
\textsuperscript{3} Trn2 UltraServer adds inter-instance rings; Trn3 UltraServer uses switched all-to-all~\cite{AWSTrn2,AWSTrn3}.\par
\textsuperscript{4} UB-Mesh uses up to 4D full mesh within a pod and Clos beyond it~\cite{liao2025ub}.\par
\textsuperscript{5} MTIA 200 uses switched PCIe; MTIA 300 uses Ethernet/RoCE across 16 devices; MTIA 400 connects 72 devices~\cite{metaMTIA300ISCA2026,metaMTIAEvolution2026}.\par
\textsuperscript{6} WSE's mesh is internal to one accelerator. FCQ: fully connected quad. An em dash means the interconnect is not characterized here.
\end{minipage}
\end{table*}

\label{sec:simt}

\textbf{GPU.} The first category is GPU, including NVIDIA~\cite{nvidiaH100Paper,nvidiaRubin2026} and AMD~\cite{amdMI300XPaper,amdMI455X2026,amdCDNA5} devices. These accelerators preserve the SIMT execution model that gives GPUs broad programmability, while incorporating tensor or matrix cores to accelerate the dense linear algebra operations prevalent in modern AI workloads. Both NVIDIA and AMD architectures employ HBM to supply data to high-throughput compute units. Representative NVIDIA systems use different NVLink organizations across generations and platforms, including direct-connect and switched fabrics~\cite{nvidiaHGX2Topology}; topology is therefore reported together with the system configuration. AMD deployments likewise range from directly connected GPU baseboards to the switched UALoE fabric of the MI455X-based Helios rack. The combination of programmability, high-bandwidth memory, and high-performance scale-up communication enables GPUs to perform effectively in large-scale AI datacenters.

\textbf{NPU.} Driven by similar demands for neural-network execution, the second category adopts a more explicitly heterogeneous NPU architecture. As summarized in Table~\ref{tab:accelerator_characteristics}, representative examples include Google TPU~\cite{tpuArch,2023ISCATPUv4,tpuv5p,tpuv5e,tpuv6e,tpuv7,tpuv8}, AWS Trainium~\cite{AWSTrainuim,AWSTrainuim2,AWSTrainuim3}, Qualcomm AI 100~\cite{qualcommAI100}, Huawei Ascend~\cite{HuaweiAscend,liao2025ub}, Intel Gaudi~\cite{IntelGaudi3HC36}, Microsoft Maia~\cite{MicrosoftMaia}, and Cambricon MLU~\cite{cambriconBangCGuide}. These accelerators integrate matrix engines with vector, scalar, and specialized functional units, enabling different operators within AI models to be mapped onto hardware blocks optimized for specific tasks. Such heterogeneity can improve area and energy efficiency when operators map well to the specialized engines and those engines remain sufficiently utilized. In particular, matrix engines accelerate GEMM-intensive computations, vector units support operations such as activations and normalization, and scalar units manage control-oriented tasks. Some NPUs replicate a complete matrix--vector--scalar core and scale out across several such cores: the Cambricon MLU, for instance, exposes a multi-core neural-processor ISA and runtime rather than a tile-placement programming model, which is why we group it with the NPUs rather than the spatial-dataflow architectures. Table~\ref{tab:accelerator_characteristics} further indicates diverse scale-up interconnect topologies, including 2D or 3D torus, Boardfly, nD full mesh, direct full mesh, switched Ethernet, and switched all-to-all networks.

Many NPUs employ systolic arrays as their primary matrix-computation engines. Instead of requiring every processing element (PE) to repeatedly access operands from a centralized register file or SRAM, data streams of activations and weights are injected from boundary PEs, while interior PEs receive operands from neighboring elements and propagate partial sums across the array. This neighbor-to-neighbor data reuse significantly reduces memory-bandwidth requirements and improves energy efficiency by replacing a large fraction of costly memory accesses with localized data transfers.

\textbf{Spatial Dataflow.} The third category maps a computation graph directly onto distributed compute, memory, and communication resources. In a spatial dataflow architecture, operator placement, local memory allocation, and data movement between operators are explicitly managed by the compiler. These features are exposed to software rather than hidden behind caches or a centralized command stream. Explicit control of data movement reduces coordination overhead, minimizes memory traffic, and improves data locality. As shown in Figure~\ref{fig:accelerator_arch}(c), spatial dataflow architectures can be grouped into three substyles: PE array, reconfigurable architecture, and functional-slice streaming processors.

\emph{PE array} architectures divide the chip into many programmable PEs, each with private SRAM and connected through an on-chip network. The compiler distributes computation across the PEs and explicitly manages data transfers between them. Different PEs can therefore execute different operators or pipeline stages. Examples include Tenstorrent~\cite{vasiljevic2024blackhole}, Meta MTIA~\cite{MTIA2,metaMTIA300ISCA2026,metaMTIAEvolution2026}, Tesla Dojo~\cite{TeslaDojo}, and Graphcore~\cite{dissectingGraphcore}. As shown in Table~\ref{tab:accelerator_characteristics}, these systems support a wider range of memory technologies than GPUs and NPUs, including GDDR6, LPDDR, DDR4, and HBM. Cerebras extends this approach to a wafer-scale PE mesh~\cite{cerebras2023}. Fabric packets, called wavelets, carry data and control information and can activate tasks on a PE. This data-triggered task scheduling coexists with local instruction sequencing: each PE executes the instructions of its selected task. The dataflow mechanism therefore determines when work becomes ready.

Other spatial dataflow architectures do not use replicated general-purpose PEs. \emph{Reconfigurable} architectures, such as SambaNova~\cite{MICRO2024sambanova}, consist of configurable compute, memory, and interconnect tiles. The compiler maps operators and data dependencies directly onto the fabric. \emph{Functional-slice streaming} processors, such as Groq~\cite{Groq2020TSP,Groq2022software}, organize the chip into fixed functional units, including matrix engines, vector engines, SRAM blocks, and switches. Data moves through these units according to a compiler-defined schedule. Unlike architectures that rely on caches, speculation, or dynamic arbitration, Groq exposes computation and data movement timing to the compiler, enabling predictable execution.

\textbf{Compute-in-Memory.} Finally, the fourth category addresses the memory bottleneck most directly through compute-in-memory architectures. Representative designs include d-Matrix~\cite{dMatrixcorsair}, SK hynix AiM~\cite{AiMJSSCC}, and Samsung PIM~\cite{Samsungaquabolt}. These accelerators integrate arithmetic into SRAM arrays or DRAM modules, using digital in-SRAM MAC arrays, or MAC-based or SIMD PIM processors. The motivation is that many AI workloads are constrained not only by arithmetic throughput, but also by the energy and latency cost of repeatedly moving weights and activations between memory and compute units. By performing part of the computation inside or near memory, compute-in-memory designs can reduce data movement and improve efficiency for bandwidth-bound operations. Although this category is more specialized than GPUs, heterogeneous NPUs, and spatial dataflow accelerators, it is increasingly important as memory bandwidth becomes a dominant system constraint.

\subsection{Compute Engine}

AI accelerators employ a diverse set of compute engines to improve performance, including SIMT processors, programmable SIMD units, dedicated matrix engines, and vector and scalar execution units. These architectural choices are shaped by several key considerations. \textit{First}, accelerator designs must balance programmability and specialization. General-purpose execution units and flexible schedulers provide broader programmability and adaptability across workloads, whereas specialized compute engines achieve higher computational efficiency, delivering greater performance and improved FLOPS per mm\textsuperscript{2}. \textit{Second}, target kernels vary in computational intensity, that is, the amount of computation performed per byte of data transferred across each memory hierarchy boundary. These differences determine the appropriate balance between compute throughput and memory bandwidth required for efficient execution. The operational characteristics of target workloads, including the different demands of AI training and inference, further influence the selection and composition of compute engines.

Graphics processing units (GPUs) were originally developed to execute gaming and graphics workloads characterized by a high degree of parallelism. Over time, GPUs evolved into highly parallel, multithreaded processors with substantial computational capability and memory bandwidth. This evolution marked the emergence of general-purpose computing on GPUs (GPGPU) and their widespread adoption for accelerating deep learning workloads~\cite{krizhevsky2012alexnet}.

Early GPU architectures were primarily dominated by SIMT processors. To improve performance on AI workloads, modern GPUs incorporate dedicated matrix engines, such as NVIDIA’s Tensor Cores, which are accessed through specialized PTX instructions. NVIDIA further supports these capabilities through high-level libraries, including cuDNN and cuBLAS, as well as templated programming abstractions such as CUTLASS and CuTe for fine-grained performance optimization. In addition, specialized functional units are introduced to accelerate transcendental operations. For instance, the B300 GPU doubles exponential operation throughput compared to B200, reflecting the heavy use of such functions in Transformer attention mechanisms~\cite{blackwell}.

NPUs are designed as domain-specific accelerators for AI workloads and therefore integrate dedicated matrix engines to accelerate GEMM-intensive computations. To support non-GEMM operations, these architectures also incorporate vector processors or specialized functional units. For example, Google TPU architectures employ systolic arrays for GEMM acceleration and integrate SparseCores to accelerate embedding operations in recommendation models~\cite{2023ISCATPUv4}. Systolic arrays are widely adopted in AI accelerators because they efficiently exploit data reuse across processing elements (PEs) within the array. As a spatial computing architecture, a systolic array reduces off-chip memory traffic by propagating data directly between neighboring PEs. To improve architectural flexibility, some NPUs additionally incorporate programmable SIMD units, thereby increasing generality and enabling support for a broader range of AI workloads and evolving model architectures, as demonstrated in the second-generation AWS NeuronCore~\cite{NeuronCorev2}.

Some architectures rely exclusively on SIMD or vector units without incorporating dedicated matrix engines, such as Cerebras systems and processing-in-memory (PIM) designs. Cerebras reports that its SRAM-centric architecture supports high utilization across selected basic linear algebra subroutines (BLAS), including low-arithmetic-intensity BLAS-1/BLAS-2 kernels that are often memory-bound on HBM-based systems~\cite{cerebras2023}. Similarly, PIM architectures developed by Samsung and SK hynix achieve high effective memory bandwidth by integrating vector or SIMD compute units in close proximity to DRAM banks~\cite{AiMJSSCC, Samsungaquabolt}.

\subsection{Memory Hierarchy}

\label{sec:memory}

\begin{figure}[ht]
    \centering
    \includegraphics[width=\linewidth]{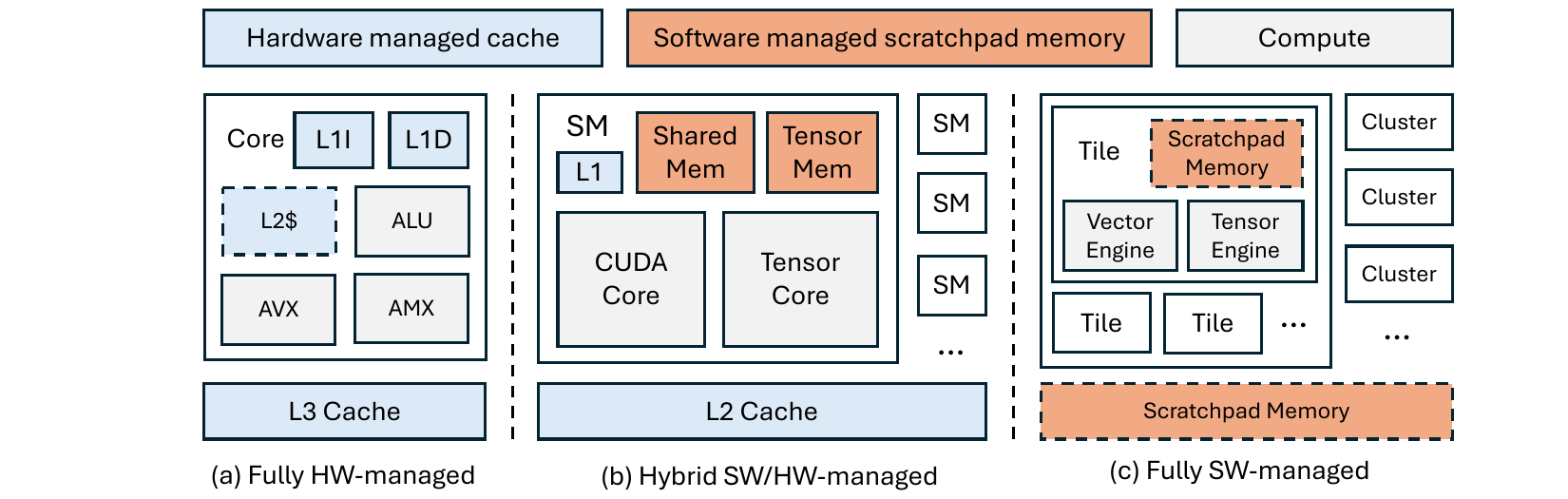}
    \caption{Hardware-managed caches and software-managed on-chip SRAM across CPUs, GPUs, and domain-specific accelerators.}
    \Description{Three panels contrast memory management. The CPU-style panel places instruction and data caches within a core above a shared last-level cache. The GPU-style panel combines hardware-managed caches with software-managed shared and tensor memory inside streaming multiprocessors. The accelerator-style panel connects vector and tensor engines to software-managed scratchpads at tile and cluster levels. Blue denotes caches, orange denotes scratchpads, and gray denotes compute.}
    \label{fig:memory}
\end{figure}

Across CPUs, GPUs, and domain-specific accelerators, on-chip SRAM spans a spectrum ranging from implicit, hardware-managed caches to explicit, software-managed scratchpad memories, as illustrated in Figure~\ref{fig:memory}. These design choices trade programmability and reduced compiler complexity against performance predictability and quality of service (QoS). CPUs lie at the hardware-managed end of this spectrum (Figure~\ref{fig:memory}(a)), relying on multi-level cache hierarchies in which the hardware determines data placement, replacement, and coherence. This implicit management simplifies programming by allowing software to operate under the abstraction of a flat address space. While such an approach offers strong generality, particularly for irregular or data-dependent access patterns, it also complicates performance and latency analysis, since cache hits, misses, and contention are governed by dynamic execution behavior.

A fully hardware-managed cache system remains relevant in today's datacenter AI ASICs. Intel's Gaudi 3, for example, features configurable SRAM that can operate as either a globally shared L3 cache or as localized L2 caches within each deep learning core. To bridge the gap between hardware transparency and the deterministic memory demands of AI workloads, Intel employs an advanced semantics-aware caching mechanism. Memory Context ID (MCID) is used to tag cache lines with shared algorithmic usage, while developers can use allocation hints to specify whether data should be cached in L2, L3, or both~\cite{IntelGaudi3HC36}.

Nevertheless, GPUs and domain-specific architectures (DSAs) increasingly expose at least portions of on-chip SRAM as explicitly software-managed scratchpads to maximize data reuse. In NVIDIA GPUs, the SRAM within each streaming multiprocessor (SM) is configurable and partitioned into a hardware-managed L1 cache and a software-managed shared memory. Through CUDA, developers explicitly allocate and orchestrate this shared memory for techniques like tiling, while still relying on conventional hardware caching (L1 and L2) for access patterns better served by hardware, as shown in Figure~\ref{fig:memory}(b). Furthermore, the latest Blackwell architecture expands this software-managed paradigm by introducing a dedicated tensor memory within each SM for tensor operations, with explicit software-managed transfers~\cite{blackwell}.

The explicit scratchpad approach shifts greater responsibility to the programmer and the compiler: tile sizing, double buffering, prefetch scheduling, and synchronization become integral to the performance contract. The benefit, however, is increased control. Data placement and movement are more deterministic, aligning well with requirements for predictable tail latency, interference isolation, and quality of service (QoS). AWS NeuronCore exemplifies this compiler-managed locality model (Figure~\ref{fig:memory}(c)), in which software-managed on-chip memory is used to maximize data locality and prefetch efficiency~\cite{NeuronCorev4}.

Another emerging trend is that memory management increasingly extends beyond the chip. Modern AI systems elevate shared-memory abstractions to the fabric level, enabling programmers or compilers to treat remote memory as part of a global address space rather than relying on explicit message passing. NVSHMEM provides GPUs with a distributed shared memory abstraction: although each GPU maintains local memory, NVSHMEM exposes a symmetric memory region on every device, allowing kernels to directly read from and write to remote GPU memory and to perform atomic operations over NVLink, PCIe, or InfiniBand~\cite{nvshmem}.

Groq further pushes the notion of shared memory over the scale-out network. It exposes a logically shared global address space backed by distributed on-chip SRAM across TSP devices~\cite{Groq2022software}. Instead of relying on adaptive, congestion-aware flow control, Groq adopts software-scheduled networking (SSN), in which the compiler resolves contention, prohibits hardware backpressure, and schedules fine-grained vector flits on each link. This approach removes the overhead associated with conventional per-flit packetization, with packet-encoding overhead of approximately 2.5\% (8 bytes per 320-byte vector). Consequently, communication behavior becomes highly deterministic, enabling stronger quality-of-service guarantees by eliminating latency variability caused by dynamic arbitration and buffering.

Computing-in-memory (CIM) offers a complementary alternative to traditional multi-level caches and scratchpad memories by performing simple arithmetic operations directly at the location where data resides, rather than staging data in separate compute arrays. SRAM-based CIM augments SRAM bit cells and peripheral circuitry to enable computation within SRAM arrays~\cite{dMatrixcorsair}. DRAM-based near- or in-memory processing integrates vector or SIMD units adjacent to DRAM banks and operates directly on data stored in row buffers, enabling bulk computations without transferring operands across the external memory interface~\cite{AimISSCC, Samsungaquabolt}. By collapsing the conventional load-compute-store sequence into localized, highly parallel operations that exploit the intrinsic bandwidth of SRAM arrays or DRAM banks, CIM can substantially reduce data movement overhead.

CIM is most effective for bandwidth-bound workloads with limited temporal reuse, such as GEMV or matrix-vector operations at small batch sizes, embedding or projection layers, and sparse or irregular computations where conventional systolic arrays cannot fully exploit data reuse. In these scenarios, colocating computation with data minimizes off-chip traffic even when reuse opportunities are limited. In contrast, workloads with high arithmetic intensity, such as large-batch GEMM, are better served by dedicated matrix engines that are explicitly designed to amplify data reuse across processing elements. Architecturally, CIM typically provides high internal parallelism but limited flexibility. Its programmability relies heavily on careful data layout and compiler scheduling to ensure operands remain resident in memory.

\section{Scale-up Interconnect}
\label{sec:scaleup}


To meet the high demands for computation and memory capacity, AI datacenters deploy clusters comprising hundreds of thousands of accelerators, integrating scale-up fabrics with Ethernet or InfiniBand scale-out networks. The scale-up fabric tightly couples accelerators within a server, rack, or pod and is designed for extremely high bandwidth and low latency. In contrast, the scale-out network, typically based on Ethernet or InfiniBand, interconnects these domains across the broader cluster. Because collective communication and model-parallel data movement are often constrained by intra-cluster communication efficiency, the scale-up fabric exerts a first-order influence on end-to-end training and inference performance. Figures~\ref{fig:scale-up-node}--\ref{fig:scale-up-pod} illustrate scale-up topologies across three tiers: the node-scale groups, typically confined to a single server with 4--8 accelerators; rack-scale groups such as NVL72, comprising multiple servers and aggregating 64--256 accelerators; larger pod (or SuperPod) fabrics, encompassing multiple racks and scaling to thousands of accelerators.

\subsection{Node Scale}

\begin{figure}[ht]
    \centering
    \includegraphics[width=\linewidth]{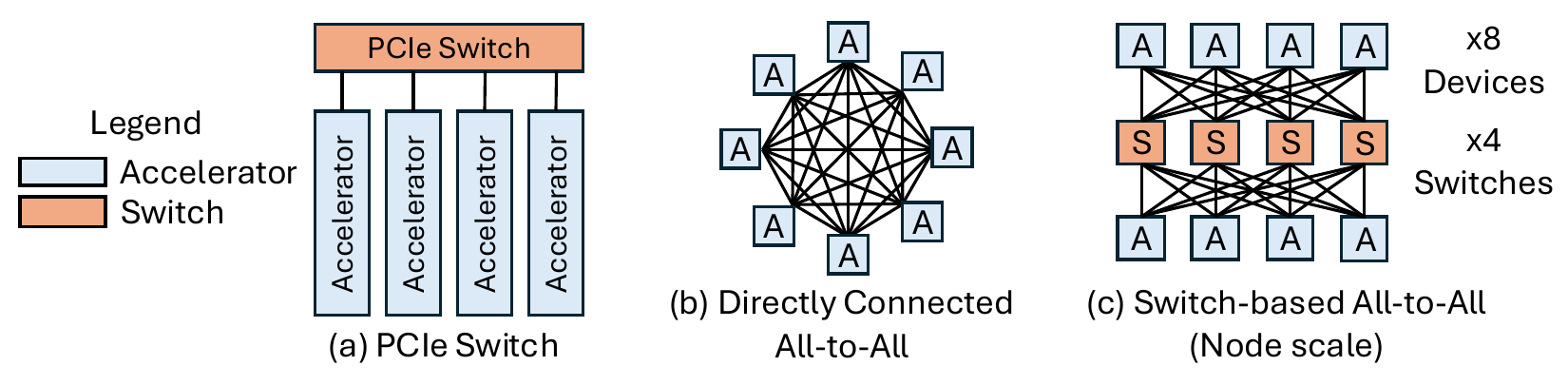}
    \caption{Node-scale interconnect organizations. (a) PCIe-switched attachment behind a host/root complex. (b) An eight-accelerator physical complete graph. (c) A dedicated switched fabric providing logical any-to-any reachability.}
    \label{fig:scale-up-node}
    \Description{Three node-scale schematics use blue boxes for accelerators and orange boxes for switches. Four accelerators attach to one PCIe switch. A second panel connects eight accelerators directly as a complete graph. A third panel connects eight accelerators through four shared switches, providing switched paths between all device pairs.}
\end{figure}

A baseline scale-up design employs one or more PCIe switches to provide packet-switched, any-to-any connectivity among CPUs, accelerators, NICs, and other peripherals, as illustrated in Figure~\ref{fig:scale-up-node}(a). This approach remains attractive due to its standards-based nature, flexibility, and relative ease of integration; however, its bandwidth is limited compared with purpose-built accelerator fabrics. Even PCIe 6.0 delivers only about 128~GB/s per direction on a $\times16$ link, which can become a bottleneck for communication-intensive workloads dominated by collectives such as AllReduce, ReduceScatter, and AllGather.

To mitigate the PCIe switch bottleneck, some systems adopt directly connected all-to-all topologies, in which accelerators are interconnected without an intermediate switch, as illustrated in Figure~\ref{fig:scale-up-node}(b). Intel’s Gaudi~3 is an explicit example of an 8-accelerator direct all-to-all fabric~\cite{IntelGaudi3HC36}. AMD Instinct platforms similarly use dense direct Infinity Fabric connectivity across eight GPUs~\cite{amdMI300XPaper}. Such direct-connect designs provide very high local bandwidth with single-hop communication, but their wiring and packaging complexity grows rapidly with system size.

To further enhance scalability, an alternative all-to-all topology incorporates dedicated scale-up switching fabrics, as illustrated in Figure~\ref{fig:scale-up-node}(c). For example, the NVIDIA HGX H100 platform employs four NVSwitch chips~\cite{nvswitch} to fully interconnect eight GPUs, and subsequent NVLink generations continue to increase per-GPU fabric bandwidth.

\subsection{Rack Scale}

\begin{figure}[ht]
    \centering
    \includegraphics[width=\linewidth]{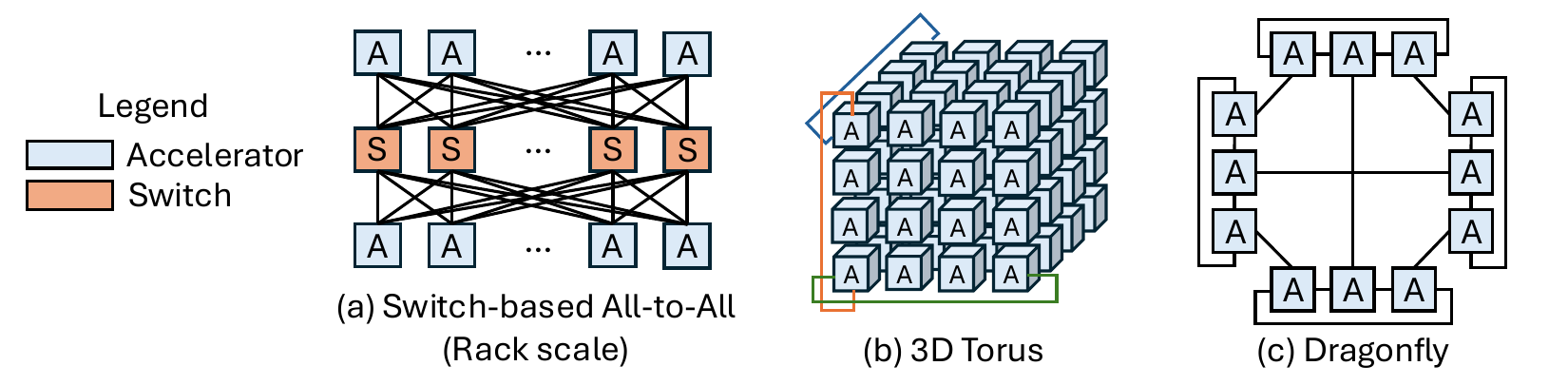}
    \caption{Rack-scale interconnect organizations. (a) A switched rack fabric, illustrated by the NVL72 organization. (b) A generic $4\times4\times4$ 3D torus with wraparound links. (c) A Dragonfly hierarchy with local and global router links.}
    \label{fig:scale-up-rack}
    \Description{Three rack-scale schematics show accelerators sharing a bank of switches, a four-by-four-by-four accelerator grid with wraparound links along three dimensions, and a Dragonfly organization with local groups joined by global links. Blue boxes denote accelerators; orange boxes denote switches where shown explicitly.}
\end{figure}

Switch-based all-to-all interconnects can be extended from the node to the rack level, as shown in Figure~\ref{fig:scale-up-rack}(a). The NVL72 rack, for instance, integrates 72 GPUs and 18 NVSwitches, organized into 18 compute trays with four GPUs and two Grace CPUs per tray and nine switch trays with two NVSwitches per tray~\cite{nvidiaNVL72Components}. AWS Trainium~3~\cite{AWSTrn3} follows a similar architectural direction with NeuronSwitch-v1, providing an all-to-all fabric within a Trn3 UltraServer containing either 64 or 144 Trainium~3 chips. Compared with direct-connect designs, switch-based all-to-all topologies scale to larger local domains and simplify board-level wiring, but they incur additional overhead in switch silicon, power consumption, and packaging.

Providing high-bandwidth any-to-any connectivity requires increasing link or switch resources as a system grows. Many architectures instead bound the per-accelerator degree and accept a larger network diameter. The torus topology, illustrated in Figure~\ref{fig:scale-up-rack}(b), is a representative example. AWS Trn1~\cite{AWSTrn1} employs a 2D torus within a single instance, while Trn2~\cite{AWSTrn2} retains a 2D torus within each 16-chip instance and connects four such instances into a 64-chip UltraServer via additional ring links. Google TPU v4~\cite{2023ISCATPUv4}, by contrast, adopts three-dimensional nearest-neighbor connectivity and can be configured as a 3D torus on supported slices. Each board integrates four TPUs, and a rack comprises 64 TPUs arranged in a $4\times4\times4$ mesh. Optical circuit switches (OCS) connect these groups and provide wraparound paths for supported torus slices, enabling reconfiguration and flexible allocation. These designs trade uniform all-to-all connectivity for bounded radix and improved scalability.

Another approach to scaling tightly coupled domains is the use of hierarchical high-radix topologies such as Dragonfly, shown in Figure~\ref{fig:scale-up-rack}(c). Rather than fully interconnecting all endpoints, Dragonfly organizes them into groups with dense local connectivity and a limited number of global links between groups, thereby reducing network diameter without incurring the quadratic wiring cost of a full mesh. Groq's earlier rack-scale work~\cite{Groq2022software} describes Dragonfly-based expansion. NVIDIA separately describes the 256-LPU Groq 3 LPX rack~\cite{NVIDIA-groq-3} as using direct chip-to-chip links within trays and a chip-to-chip spine across trays, without identifying that rack as Dragonfly.

\subsection{Pod Scale}

\begin{figure}[ht]
    \centering
    \includegraphics[width=0.8\linewidth]{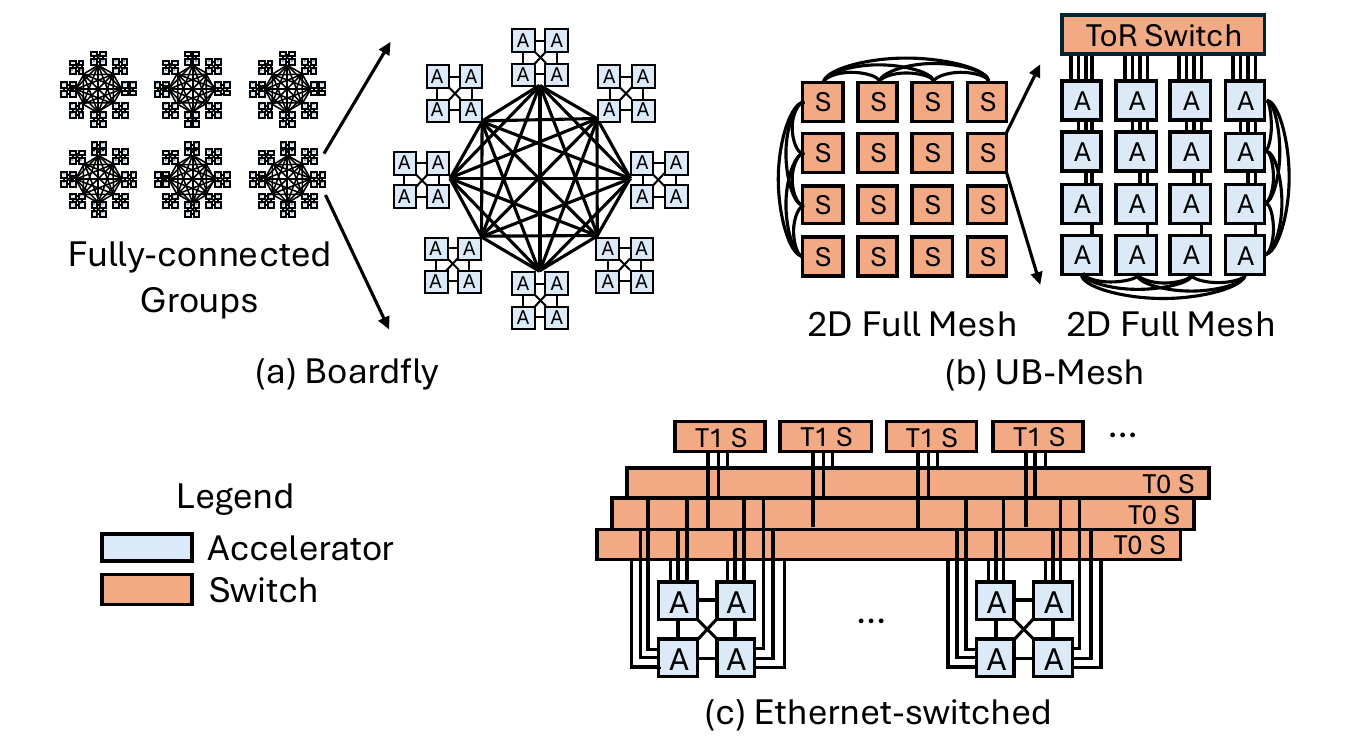}
    \caption{Pod-scale interconnect organizations. (a) The announced Boardfly hierarchy. (b) The proposed UB-Mesh hierarchy. (c) The Maia 200 hierarchy combining local fully connected quads with Ethernet switching.}
    \label{fig:scale-up-pod}
    \Description{Three hierarchical pod networks are shown. Boardfly expands a fully connected group into interconnected four-accelerator boards. UB-Mesh shows a two-dimensional full mesh of switches alongside a two-dimensional accelerator mesh connected to a top-of-rack switch. The Ethernet-switched panel connects local fully connected accelerator quads to tier-zero and tier-one switches.}
\end{figure}

Vendors are increasingly exploring deeper hierarchical designs at the pod scale. Google's forthcoming TPU~8i introduces Boardfly~\cite{tpuv8}, a topology inspired by Dragonfly that aggregates four-chip building blocks into eight-board groups and interconnects 36 such groups via optical circuit switches (Figure~\ref{fig:scale-up-pod}(a)). Unified scale-up and scale-out network topologies are also deployed at the pod scale, such as Huawei’s UB-Mesh~\cite{liao2025ub} and Microsoft's Maia~\cite{MicrosoftMaia, dighe2026maia200}.

Huawei’s UB-Mesh~\cite{liao2025ub} adopts a different strategy, proposing a hierarchically localized nD-FullMesh built on a unified bus and protocol stack. Its published pod-scale design realizes a 4D FullMesh by composing full-mesh connectivity both within and across racks, supported by dedicated low-radix and high-radix switches (Figure~\ref{fig:scale-up-pod}(b)). Microsoft Maia adopts an Ethernet-based alternative in scale-up fabrics. As illustrated in Figure~\ref{fig:scale-up-pod}(c), Maia uses a two-tier Ethernet-switched topology, where the first tier forms a switchless fully connected quad (FCQ), in which four Maia accelerators are directly connected; the second tier uses Ethernet switches to extend the scale-up domain across racks to as many as 6,144 accelerators. 


Viewed through the topology trade-offs discussed above, PCIe switching provides a broadly supported baseline, although the surveyed deployments generally offer lower per-accelerator bandwidth than purpose-built scale-up fabrics. Direct full meshes provide single-hop communication, but their link count grows quadratically with the number of endpoints. Switched any-to-any fabrics can accommodate concurrent collectives and irregular traffic when switch capacity, routing, and endpoint injection bandwidth are sufficient, at the cost of additional switch power and packaging complexity. Tori preserve bounded node degree as the system scales, but their larger diameter can increase communication latency. Dragonfly- and Boardfly-like hierarchies reduce path length without requiring full-mesh connectivity at system scale, while UB-Mesh and Ethernet-switched designs introduce additional tiers of local and global connectivity. Collectively, these approaches illustrate how topology is co-designed across node-, rack-, and pod-scale systems to balance bandwidth, latency, scalability, and implementation cost.


\subsection{Collective Communications}

\begin{figure}[htbp]
    \centering
    \includegraphics[width=0.9\textwidth]{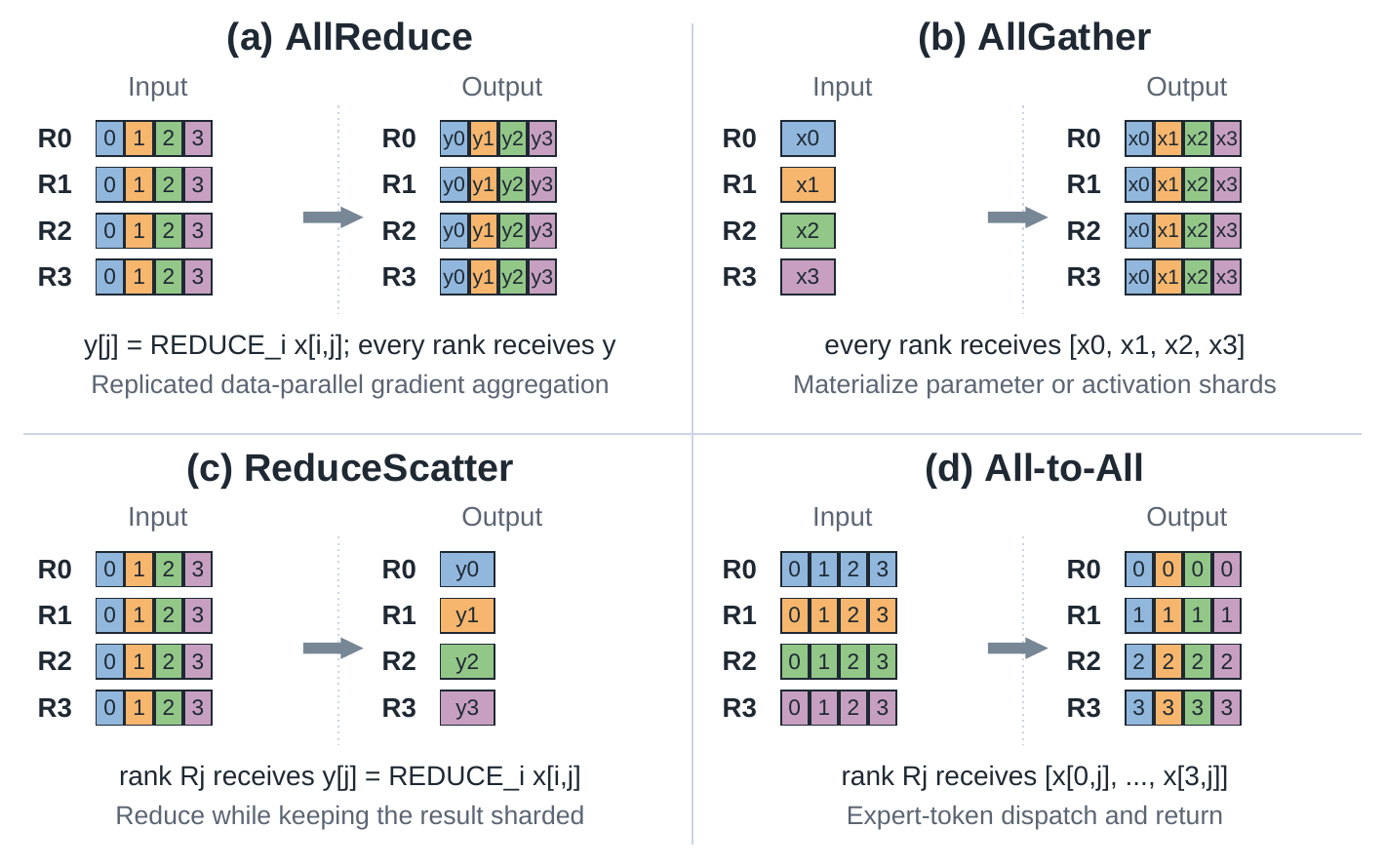}
    \caption{Endpoint semantics of four common collectives for ranks R0--R3. (a) AllReduce returns the elementwise reduction of all inputs to every rank. (b) AllGather concatenates the rank-local shards at every rank. (c) ReduceScatter reduces all inputs and retains one result shard per rank. (d) All-to-All sends a distinct destination chunk to each rank. The panels describe data transformations, not communication routes, algorithms, or physical topologies.}
    \label{fig:collective-communications}
    \Description{Each panel shows input and output blocks at four ranks, R0 through R3. AllReduce leaves the same four reduced blocks at every rank. AllGather copies each rank's single input shard into an ordered four-shard output at every rank. ReduceScatter leaves reduced block j only at rank j. All-to-All changes blocks grouped by source into blocks grouped by destination, so each rank receives one block from each source.}
\end{figure}

The node-, rack-, and pod-scale topologies above define the paths, locality hierarchy, and contention domains available to distributed workloads, while collective algorithms determine how communication is mapped onto those physical resources. Collective operations specify an endpoint data transformation, whereas a collective algorithm specifies the communication schedule used to realize it on a physical topology. Figure~\ref{fig:collective-communications} illustrates four operations that dominate distributed AI training. AllReduce combines corresponding elements from every rank and returns the complete reduced tensor to every rank. AllGather collects rank-local shards and returns their ordered concatenation to every rank. ReduceScatter first reduces corresponding elements and then leaves rank~$i$ with only shard~$i$ of the result. All-to-All instead transposes rank-specific chunks: every rank sends a distinct chunk to every peer and receives one chunk from each source~\cite{nvidiaNCCLCollectives}. ReduceScatter followed by AllGather can implement AllReduce, but this decomposition is an algorithmic choice rather than the definition of AllReduce.

These operations play different roles in a training stack. Replicated data parallelism typically uses AllReduce to aggregate gradients. Fully sharded data parallelism and ZeRO-style training use AllGather to materialize parameter shards before computation and ReduceScatter to aggregate gradients while preserving sharding~\cite{rajbhandari2020zero}. Tensor and sequence parallelism use combinations of AllReduce, AllGather, and ReduceScatter to exchange partial activations or tensor slices~\cite{shoeybi2019megatron,korthikanti2023sequence}. Expert-parallel mixture-of-experts models use All-to-All for token dispatch to experts and again to return expert outputs~\cite{lepikhin2021gshard}. Thus, the operation determines the required data movement, but not the best schedule.

For large reduction payloads on relatively uniform links, a ring is often attractive. It partitions the tensor and pipelines ReduceScatter and AllGather around a logical cycle, approaching bandwidth-optimal per-rank traffic~\cite{patarasuk2009ring}. Its cost, however, includes a number of startup stages linear in the rank count, so latency becomes increasingly important for small messages or very large groups. Tree algorithms reduce the number of dependent stages to logarithmic depth. Double binary trees process different halves of the tensor on complementary trees, distributing the forwarding load and, with a suitable mapping, retaining high aggregate bandwidth; they are therefore effective for latency-sensitive and medium-sized reductions at scale~\cite{jeaugey2019nccltrees}. A ring or tree is a logical schedule and need not match the physical wiring literally.

Several other algorithm families fill the space between these regimes. Recursive halving for ReduceScatter followed by recursive doubling for AllGather, often called a Rabenseifner-style AllReduce, uses logarithmic rounds and works well when the fabric can sustain concurrent partner exchanges. Bruck-style schedules similarly reduce startup rounds for small AllGather and All-to-All messages, whereas pairwise exchange is commonly preferable for bandwidth-dominated All-to-All traffic~\cite{thakur2005mpich}. Parallel Aggregated Trees (PAT) extend logarithmic-step AllGather and ReduceScatter to arbitrary rank counts while limiting long-distance transfers~\cite{jeaugey2025pat}. For irregular or asymmetric systems, TACCL demonstrates that topology-specific schedules can instead be synthesized from a hardware model and a communication sketch~\cite{shah2023taccl}.

The physical topology then determines which logical schedule can use the available links without congestion. Direct and switched any-to-any fabrics in Figures~\ref{fig:scale-up-node}(b)--(c) and~\ref{fig:scale-up-rack}(a) can support multiple concurrent rings or trees; their direct reachability also suits pairwise All-to-All. On bounded-degree meshes and tori, a runtime can decompose a collective by dimension or embed multiple rings so that traffic follows local links and avoids oversubscribing a cut~\cite{2023ISCATPUv4}. Dragonfly, Boardfly, and node--rack--pod fabrics favor hierarchical local--global--local schedules: reduce or gather within a group, exchange only the necessary data across scarce inter-group links, and distribute within the destination group. This same principle applies to the deeper locality exposed by UB-Mesh and Maia, without requiring a separate collective semantic for each fabric~\cite{liao2025ub,dighe2026maia200}. On fabrics with in-network reduction, CollNet- or NVLS-style schedules can offload the reduction phases of AllReduce or ReduceScatter, but AllGather and All-to-All still require moving their distinct data to the destinations~\cite{nvidiaNCCLAlgorithms}.

Consequently, there is no universally best collective algorithm. Practical runtimes select among rings, trees, recursive exchanges, and hierarchical schedules according to the operation, message size, number of ranks, topology, link asymmetry, and current contention~\cite{thakur2005mpich,nvidiaNCCLAlgorithms}. Performance depends on both the fabric's raw bandwidth and how well the runtime maps the operation onto its locality, path diversity, and hierarchy.

\FloatBarrier

\section{Architecture and System Evolution across Generations}

\label{sec:evolution}


The rapid scaling of AI model sizes and the continual evolution of model architectures impose new requirements on AI accelerator design. To examine these trends, we analyze the architectural evolution of AI accelerators across multiple generations of NVIDIA and AMD GPUs, as well as Google TPUs and AWS Neuron accelerators. On the compute side, accelerators increasingly adopt lower-precision arithmetic to improve performance, as reduced precision enables higher FLOPS within the same hardware resource constraints. In addition, structured sparsity has been introduced to efficiently accelerate sparse computations by skipping zero-valued operands in matrices.

As compute units become more powerful, they demand correspondingly higher data delivery rates. This requirement is reflected in advances in scheduling mechanisms, data movement support, and the continued growth of on-chip SRAM capacity. Moreover, because AI model sizes scale significantly faster than hardware FLOPS and memory capacity (Figure~\ref{fig:scaling}), modern AI datacenters increasingly rely on aggregating large numbers of accelerators. To meet the resulting inter-device communication demands, accelerator designs have substantially increased interconnect bandwidth and adopted interconnect topologies that evolve in tandem with emerging AI model architectures.

\subsection{Lower Precision and Structured Sparsity}

\begin{figure}[ht]
    \centering
    \includegraphics[width=0.9\linewidth]{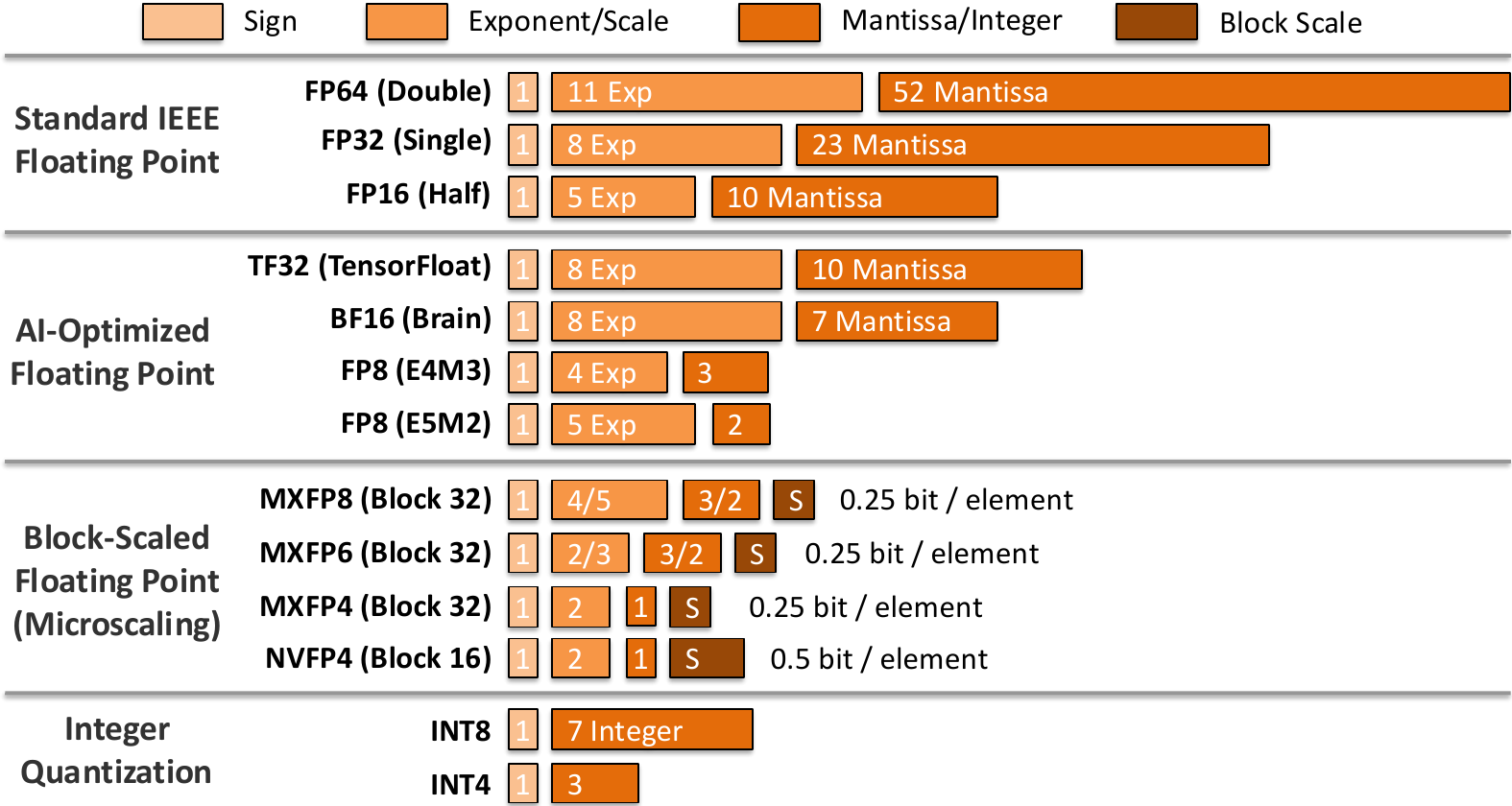}
    \caption{Bit allocation of standard, AI-oriented, block-scaled, and integer formats. For block-scaled formats the shared 8-bit scale is drawn as its amortized per-element cost (0.25 bit per element for 32-element blocks, 0.5 bit for 16-element blocks); tensor-level scales are omitted. MX and NVFP4 layouts follow the OCP MX specification~\cite{ocpMX2023} and NVIDIA's NVFP4 format description~\cite{nvidiaNVFP4Format2025}, respectively.}
    \label{fig:precision}
    \Description{Rows compare bit allocations for IEEE floating point, AI-oriented floating point, block-scaled formats, and integers. Sign, exponent, and mantissa lengths are 1/11/52 for FP64, 1/8/23 for FP32, 1/5/10 for FP16, 1/8/10 for TF32, and 1/8/7 for BF16. FP8 uses E4M3 or E5M2. MXFP8, MXFP6, and MXFP4 share a scale across 32 values; NVFP4 uses blocks of 16. Separate rows show INT8 and INT4.}
\end{figure}

Figure~\ref{fig:precision} presents a comparative taxonomy of bit allocations across conventional high-performance computing (HPC) and emerging AI-optimized numerical formats. These formats are grouped into four categories: standard IEEE floating-point formats (e.g., FP64 and FP32), AI-optimized floating-point formats (e.g., TF32 and BF16), block-scaled floating-point formats (microscaling), and integer quantization. The horizontal bars depict the internal composition of each format, with segments representing the sign bit, exponent or scale, mantissa or integer data, and shared block scale. This decomposition highlights the fundamental trade-offs between dynamic range, primarily determined by exponent width, and numerical precision, governed by mantissa width, as architectures evolve from general-purpose double precision toward specialized low-precision representations.

The visualization reveals two key trends in modern arithmetic design: the prioritization of dynamic range and the increasing use of amortized scaling. Within the AI-optimized category, formats such as BF16 and TF32 preserve an 8-bit exponent comparable to FP32 to maintain training stability, while substantially reducing mantissa width to lower hardware cost. Block-scaled formats, including MXFP and NVFP, share scale factors across groups of values, trading scale-storage overhead against dynamic range and quantization error.

\begin{figure}[ht]
    \centering
    \includegraphics[width=0.5\linewidth]{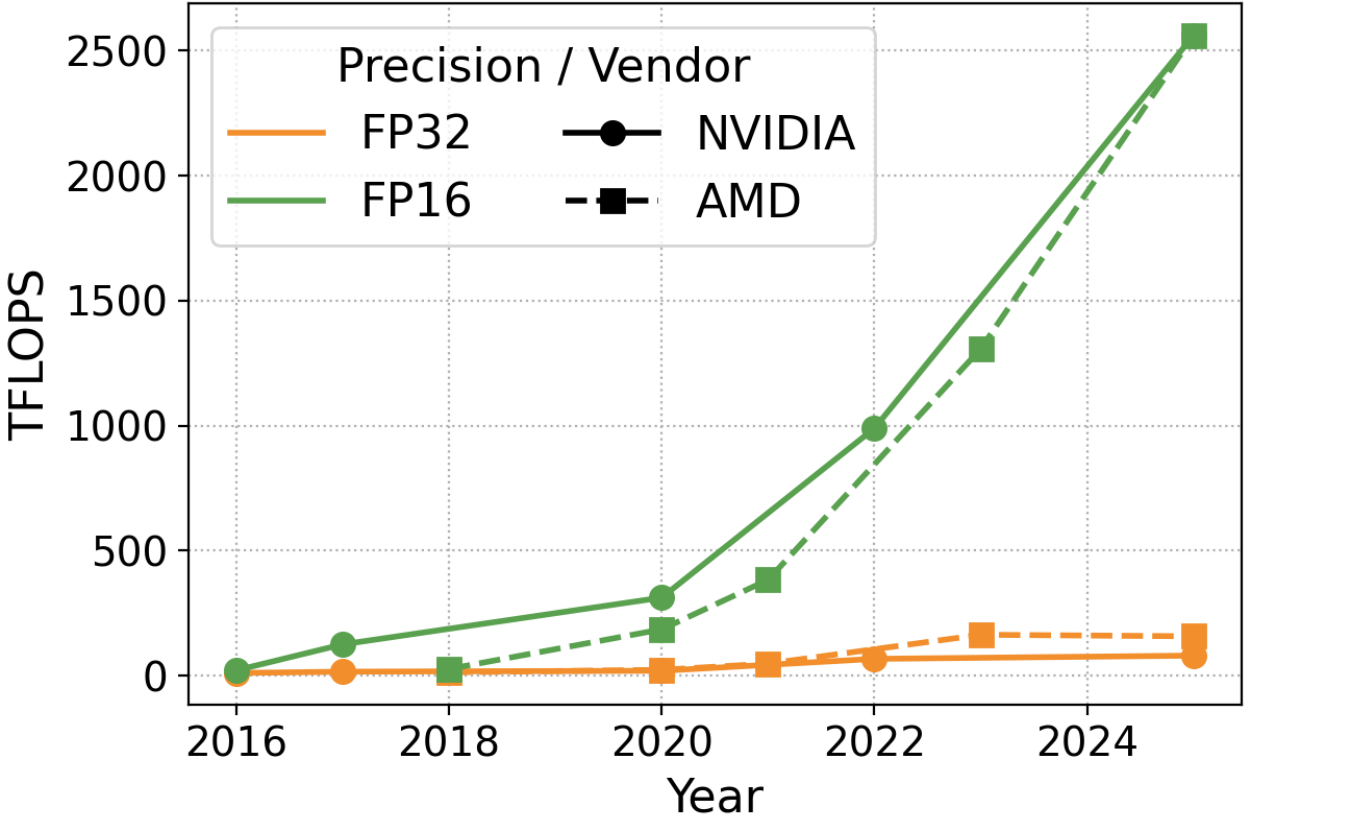}
    \caption{Per-GPU peak FP32 throughput and dense FP16/BF16 matrix throughput across NVIDIA and AMD GPUs.}
    \label{fig:fp32fp16}
    \Description{A line plot shows peak throughput in TFLOPS against year from 2016 to 2025. Green FP16 matrix-throughput curves rise much more steeply than orange FP32 curves. NVIDIA uses solid lines with circles and AMD uses dashed lines with squares. By the last observations, matrix throughput is around 2,500 TFLOPS, while FP32 throughput remains below 200 TFLOPS.}
\end{figure}

We summarize the peak computational throughput of modern AI accelerators across numerical formats in Table~\ref{tab:peak_tflops_vendor_multirow}. In GPUs, high-precision general-purpose SIMD units, such as FP64 and FP32, exhibit relatively slow scaling compared to the rapid growth of low-precision FP16 matrix engines, together with the introduction of even lower-precision formats, including FP8 and FP4. Figure~\ref{fig:fp32fp16} further compares per-GPU peak FP32 throughput with dense FP16/BF16 matrix throughput across NVIDIA and AMD GPUs.

\begin{table*}[ht]
\centering
\caption{Peak computational throughputs of representative AI accelerators.}
\resizebox{\textwidth}{!}{%
\label{tab:peak_tflops_vendor_multirow}
\begin{tabular}{l l l r r r r r r r}
\toprule
\multirow{2}{*}{Vendor} & \multirow{2}{*}{Year} & \multirow{2}{*}{Accelerator} & FP64 & FP32 & TF32 & FP16/BF16 & FP8 & INT8 & FP4 \\
& & & TFLOPS & TFLOPS & TFLOPS & TFLOPS & TFLOPS & TOPS & TFLOPS \\
\midrule

\multirow{6}{*}{NVIDIA}
 & 2016 & P100~\cite{nvidiaP100Paper}\textsuperscript{a} & 5.3 & 10.6 & -- & 21.2 & -- & -- & -- \\
 & 2017 & V100~\cite{nvidiaV100Paper} & 7.8 & 15.7 & -- & 125 & -- & -- & -- \\
 & 2020 & A100~\cite{nvidiaA100Paper} & 9.7 & 19.5 & 156 & 312 & -- & 624 & -- \\
 & 2022 & H100~\cite{nvidiaH100Paper} & 34 & 67 & 495 & 989 & 1979 & 1979 & -- \\
 & 2025 & B300~\cite{B300WhitePaper} & 1.3 & 80 & 1250 & 2500 & 5000 & 165\textsuperscript{d} & 15000 \\
 & 2026 & Rubin GPU~\cite{nvidiaRubin2026} & 33 & 130 & 2000 & 4000 & 17500 & -- & 35000 \\
\midrule

\multirow{6}{*}{AMD GPU}
 & 2018 & MI50~\cite{amdMI50Specs} & 6.6 & 13.3 & -- & 26.5 & -- & 53 & -- \\
 & 2020 & MI100~\cite{MI100WhitePaper,amdMI100Brief} & 11.5 & 23.1 & -- & 184.6\textsuperscript{b} & -- & 184.6 & -- \\
 & 2021 & MI250X~\cite{MI250WhitePaper} & 47.9 & 47.9 & -- & 383 & -- & 383 & -- \\
 & 2023 & MI300X~\cite{MI300XWhitePaper} & 81.7 & 163.4 & 653.7 & 1307 & 2614 & 2614 & -- \\
 & 2025 & MI355X~\cite{MI355XWhitePaper,amdMI355XDatasheet} & 78.6 & 157.3 & -- & 2517 & 5033 & 5033 & 10066 \\
 & 2026 & MI455X~\cite{amdMI455X2026,amdCDNA5} & 5 & 315 & -- & 5000 & 20100 & 5000 & 40300 \\
\midrule
\multirow{7}{*}{Google TPU}
 & 2021 & TPU~v4~\cite{2023ISCATPUv4} & -- & -- & -- & 275 & -- & 275 & -- \\
 & 2023 & TPU~v5e~\cite{tpuv5e} & -- & -- & -- & 197 & -- & 393 & -- \\
 & 2023 & TPU~v5p~\cite{tpuv5p} & -- & -- & -- & 459 & 459\textsuperscript{c} & -- & -- \\
 & 2024 & TPU~v6e~\cite{tpuv6e,googleTPUComparison2026} & -- & -- & -- & 918 & 918\textsuperscript{c} & 1836 & -- \\
 & 2025 & TPU~v7~\cite{tpuv7} & -- & -- & -- & 2307 & 4614 & -- & -- \\
 & 2026 & TPU~8t~\cite{tpuv8} & -- & -- & -- & -- & -- & -- & 12600 \\
 & 2026 & TPU~8i~\cite{tpuv8} & -- & -- & -- & -- & -- & -- & 10100 \\
\midrule

\multirow{5}{*}{AWS Neuron}
 & 2018 & Inferentia 1~\cite{AWSInferentia} & -- & -- & -- & 64 & -- & 128 & -- \\
 & 2020 & Trainium 1~\cite{AWSTrainuim} & -- & 47.5 & 190 & 190 & 190 & 380 & -- \\
 & 2022 & Inferentia 2~\cite{AWSInferentia2} & -- & 47.5 & 190 & 190 & 190 & 380 & -- \\
 & 2023 & Trainium 2~\cite{AWSTrainuim2} & -- & 181 & 667 & 667 & 1299 & -- & -- \\
 & 2024 & Trainium 3~\cite{AWSTrainuim3} & -- & 183 & 671 & 671 & 2517 & -- & 2500 \\
\bottomrule
\end{tabular}%
}
\vspace{2pt}
\begin{minipage}{\textwidth}
\vspace{1mm}
\scriptsize
Vendor-reported dense peaks in decimal units; a fused multiply-add counts as two operations. FP32/FP64 use the listed scalar/vector peaks, and lower precisions use matrix engines where available. Years denote first public family disclosure. ``--'' means not specified by the cited source or unsupported.\par
\textsuperscript{a} P100 FP16 uses CUDA cores; P100/V100 values do not imply BF16 support.\par
\textsuperscript{b} MI100 achieves 184.6 TFLOPS FP16 but 92.3 TFLOPS BF16. TPU/Neuron entries use BF16.\par
\textsuperscript{c} Google's reported FP8 peaks for v5p/v6e equal their BF16 peaks; these should not be interpreted as a native FP8 throughput increase. TPU 8t/8i dashes indicate unreported values in the cited announcement.\par
\textsuperscript{d} B300 denotes the GB300 NVL72 GPU: the cited datasheet reports 330 sparse INT8 TOPS and specifies dense throughput as half. MI455X's 5 TFLOPS FP64 and 5 POPS INT8 follow AMD's specification; Trainium 3 processes MXFP4 through its MXFP8 path.
\end{minipage}
\end{table*}

Modern AI accelerators also exploit structured sparsity to improve performance and efficiency by systematically zeroing less important weights. Empirically, trained networks contain substantial redundancy: after pruning or during inference, many weights can be set to zero with little accuracy loss. Leveraging these zeros reduces both computation and memory traffic. While fully unstructured sparsity is difficult for hardware due to irregular access and heavy metadata overhead, structured or semi-structured sparsity enforces regular patterns, typically an $M:N$ rule in which only $M$ of every $N$ values are nonzero. This regularity preserves locality, enables efficient compression and fetching, and remains compatible with matrix engines~\cite{sparseDNN}.

\begin{figure}[ht]
    \centering
    \includegraphics[width=0.6\linewidth]{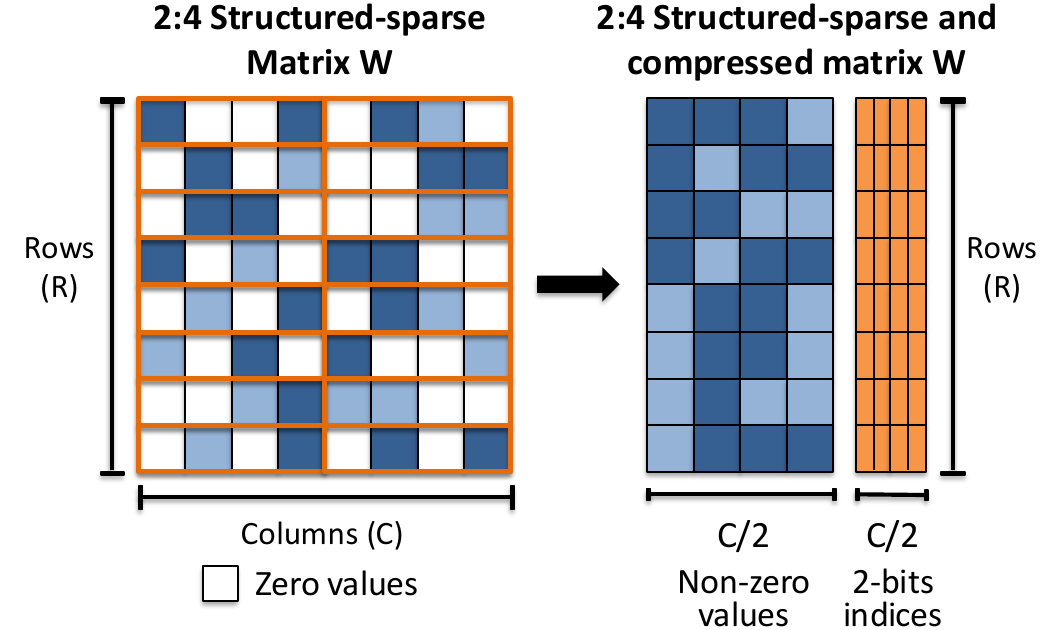}
    \caption{Example of 2:4 structured-sparse matrix representation}
    \label{fig:sparsity}
    \Description{A matrix with R rows and C columns contains two nonzero entries in every group of four entries along a row. An arrow maps it to an R-by-C/2 array containing only nonzero values and a separate R-by-C/2 array of two-bit indices recording their original positions. White input cells denote zeros.}
\end{figure}

Accelerators realize these benefits by extending tensor or matrix engines with sparsity-aware formats that skip multiplications on zeros without disrupting a regular datapath. NVIDIA and AMD GPUs primarily support a fixed 2:4 pattern, where two of every four elements are zero. An example of 2:4 sparsity is illustrated in Figure~\ref{fig:sparsity}. Operands are stored in compact form with lightweight metadata, enabling up to $\sim$2$\times$ higher throughput when the constraint is met~\cite{A100WhitePaper,sparsity,MI300XWhitePaper}. AWS Neuron generalizes this approach with flexible $M:N$ patterns, such as 4:16, 4:12, 4:8, 2:8, 2:4, 1:4, and 1:2, allowing tunable accuracy–throughput trade-offs~\cite{NeuronCorev4}.

\subsection{Data Path and Control Flow}


As matrix-engine performance has increased, AI accelerators have correspondingly optimized their data paths and control flow to sustain the higher computational throughput. To illustrate this trend, we conduct an evolutionary case study across five generations of NVIDIA Tensor Cores from Volta to Blackwell architectures. 

\begin{figure}[ht]
    \centering
    \includegraphics[width=\linewidth]{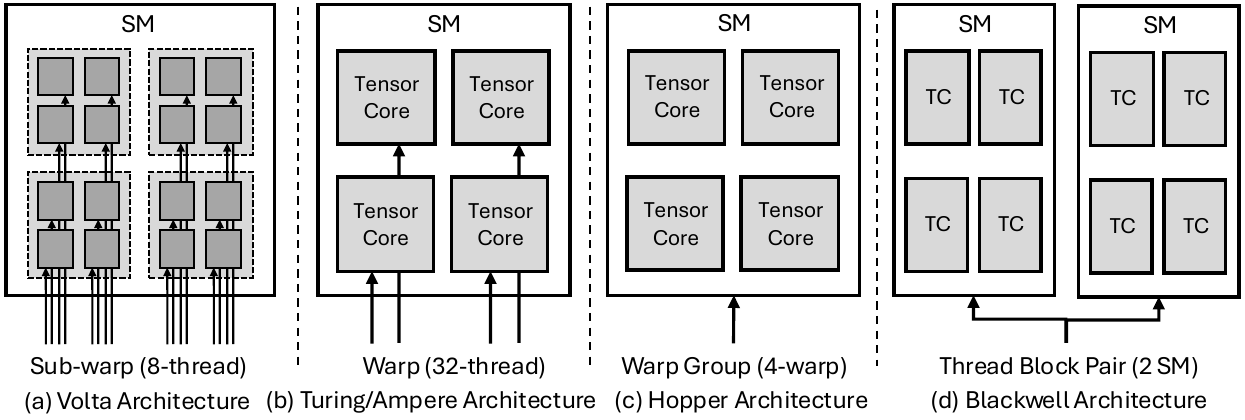}
    \caption{Tensor core control flow evolution across NVIDIA GPU architectures}
    \label{fig:GPU_control_flow}
    \Description{Four panels show increasing tensor-operation coordination scope: an eight-thread sub-warp in Volta, a 32-thread warp in Turing and Ampere, a four-warp group in Hopper, and a thread-block pair spanning two streaming multiprocessors in Blackwell. Arrows connect each coordinating thread group to the tensor cores involved.}
\end{figure}

NVIDIA GPUs have progressively redesigned both data-feeding mechanisms and control granularity across Tensor Core generations. The data path has evolved from conventional per-thread \texttt{ld} instructions to warp-level \texttt{wmma.load}, and further to more flexible \texttt{ldmatrix} operations involving groups of four threads, while introducing asynchronous data movement to better overlap data transfer with computation. In parallel, the control granularity of Tensor Core execution has expanded from sub-warp groups (8-thread) to full warp (32-thread), warp groups (4-warp), and, more recently, coordinated execution across multiple streaming multiprocessors (SM), as illustrated in Figure~\ref{fig:GPU_control_flow}.

\textbf{Volta: sub-warp group control granularity.}
The first-generation Tensor Cores are exposed as warp-cooperative at the programming level, but their machine-level execution granularity is finer. Each 32-thread warp is executed on Tensor Cores as four 8-thread \texttt{quadpair}-issued micro-operations~\cite{CuTeMMA}. Volta also introduces the \texttt{wmma.load} instruction to efficiently feed data to Tensor Cores by loading $N \times 128$ bytes of data from shared memory into registers.

\textbf{Turing: full-warp control granularity and flexible data loading.}
Turing shifts instruction issuing from 8-thread \texttt{quadpair} groups to the full-warp level, distributing operand fragments across all 32 lanes and executing Tensor Core instructions at warp granularity~\cite{yan2020demystifying}. It also introduces \texttt{ldmatrix}, which uses a group of four threads to load 16 consecutive bytes, offering higher efficiency than per-thread 4-byte \texttt{ld} operations and better flexibility than earlier warp-level 128-byte granularity of \texttt{wmma.load}. The key advantage is \emph{layout flexibility}: kernels can apply shared-memory permutations, such as CUTLASS-style swizzles, to reduce bank conflicts. An ablation study shows that replacing a baseline matrix multiply–accumulate (\texttt{MMA}) kernel with a permuted shared-memory layout enabled by \texttt{ldmatrix} achieves up to a $3\times$ speedup over \texttt{wmma.load}~\cite{sun2022dissecting}.

\begin{figure}[ht]
    \centering
    \includegraphics[width=0.9\linewidth]{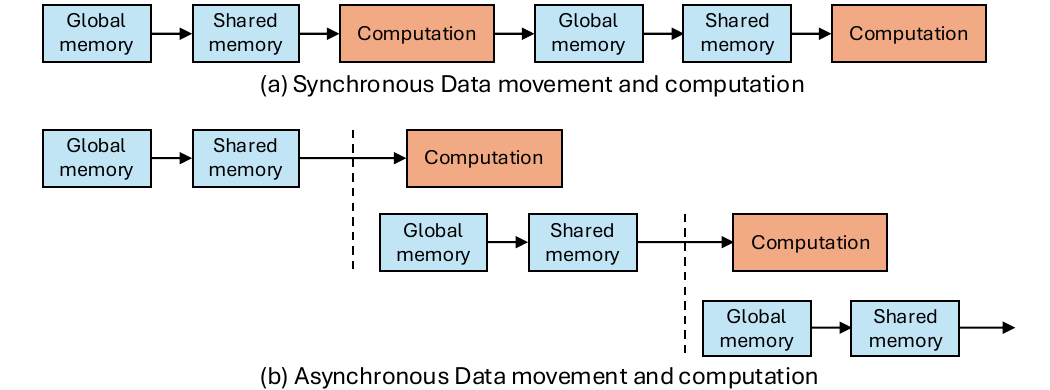}
    \caption{Comparison between synchronous and asynchronous data paths}
    \label{fig:GPU_data_path}
    \Description{The upper pipeline repeats global-memory access, shared-memory staging, and computation in sequence. The lower pipeline staggers successive iterations so that global-to-shared-memory transfers for the next iteration overlap computation on the current one. Dashed vertical lines separate stages in the overlapping schedule.}
\end{figure}

\textbf{Ampere: hiding global-memory latency with asynchronous copy pipelines.}
As Tensor Core compute continued to scale, shared-memory tiling alone became insufficient when global-to-shared data movement serialized with computation. Ampere introduces \texttt{cp.async}, which transforms operand movement into a software pipeline: while Tensor Cores operate on tiles resident in shared memory, subsequent tiles are prefetched asynchronously from global memory, as illustrated in Figure~\ref{fig:GPU_data_path}. This design reduces synchronization overhead and enables effective overlap between computation and data movement. As a result, the asynchronous implementation of an \texttt{MMA} kernel achieves approximately a $2\times$ speedup over a synchronous baseline~\cite{sun2022dissecting}.

\textbf{Hopper: asynchronous warp-group control, bulk transfers, and direct SM-to-SM data movement.}
Hopper further widens the control scope and deepens the execution pipeline. The introduction of \texttt{WGMMA} instructions expands the unit of issue from a single warp to a warp group, typically comprising four warps.
This design addresses utilization limits observed with warp-level MMA on Hopper GPUs, where microbenchmarking reports performance plateaus at approximately 63\% of peak without warp-group–level pipelining~\cite{luo2024benchmarking}.

On the operand-delivery path, Hopper introduces the Tensor Memory Accelerator (TMA) to offload bulk global-to-shared transfers and their associated address-generation and loop overheads, which increasingly constrain performance as Tensor Core throughput rises. Instead of having many threads cooperatively execute numerous fine-grained copies (e.g., \texttt{cp.async}) with explicit per-thread addressing, Hopper allows a \emph{single} designated warp lane (“warp leader”) to enqueue a descriptor-driven, asynchronous tensor transfer. In a GEMM microbenchmark with $M{=}128$, $N{=}4096$, and $K{=}4096$, substituting Ampere-style \texttt{cp.async} staging with TMA increases the profiled global-memory throughput from $\sim$910~GB/s to 1.45~TB/s (a 59\% increase in that kernel's global-memory throughput)~\cite{HopperTMA}.

Hopper further introduces distributed shared memory within an SM cluster, enabling direct SM-to-SM data movement without leveraging global memory. Microbenchmark results report an SM-to-SM latency of approximately 180 cycles, compared to about 265 cycles when accessing data via L2 cache, corresponding to a latency reduction of roughly 32\%~\cite{luo2024benchmarking}.

\textbf{Blackwell: dedicated Tensor Memory (TMEM) and multi-SM tensor execution.}
Blackwell further centralizes the datapath around Tensor Cores by introducing 256~KB of Tensor Memory (TMEM) per SM. TMEM can store intermediate operands for Tensor Cores and significantly increases effective data bandwidth. In back-to-back FP8 GEMM experiments, TMEM sustains approximately 8~TB/s, about $2.1\times$ higher than \texttt{ld.global}-dominated paths that saturate near 3.8~TB/s~\cite{jarmusch2025microbenchmarking}.

Blackwell also extends the cooperation boundary beyond a single SM through SM-pair execution, in which two adjacent thread blocks are co-scheduled within an SM cluster to jointly execute a larger \texttt{MMA} kernel. By sharing operands across SMs, this mechanism reduces redundant control overhead and improves utilization. Blackwell continues the trajectory established by Hopper, emphasizing wider cooperation and increasingly asynchronous, specialized data paths to keep rapidly scaling Tensor Cores fully utilized.

\subsection{Memory Hierarchy and Scale-up Interconnect}

\begin{table*}[t]
\centering
\caption{Memory systems and scale-up interconnects of representative AI accelerators.}
\label{tab:mem_interconnect_vendor_multirow}
\resizebox{\textwidth}{!}{%
\begin{tabular}{l l l r r r r c}

\midrule
\multirow{3}{*}{Vendor} & \multirow{3}{*}{Year} & \multirow{3}{*}{Accelerator} & Memory & Memory & On-chip & Interconnect\textsuperscript{a} & \multirow{3}{*}{\shortstack{Scale-up\\Topology}} \\
& & & Capacity & Bandwidth & SRAM & Bandwidth & \\
& & & (GB) & (TB/s) & (MB; type) & (GB/s) & \\
\midrule
\multirow{6}{*}{NVIDIA GPU}
 & 2016 & P100~\cite{nvidiaP100Paper} & 16 & 0.7 & 4 & 160 & Hybrid cube mesh \\
 & 2017 & V100~\cite{nvidiaV100Paper} & 32 & 0.9 & 6 & 300 & Hybrid cube mesh \\
 & 2020 & A100~\cite{nvidiaA100Paper} & 80 & 2.0 & 40 & 600 & All-to-All \\
 & 2022 & H100~\cite{nvidiaH100Paper} & 80 & 3.4 & 50 & 900 & All-to-All \\
 & 2025 & B300~\cite{B300WhitePaper,blackwell} & 279 & 8 & 126 & 1800 & All-to-All \\
 & 2026 & Rubin~\cite{nvidiaRubin2026} & 288 & 19.2 & -- & 3000 & All-to-All \\
\midrule

\multirow{6}{*}{AMD GPU}
 & 2018 & MI50~\cite{amdMI50Specs} & 32 & 1.0 & 4 & 184 & Ring \\
 & 2020 & MI100~\cite{MI100WhitePaper,amdMI100Brief} & 32 & 1.2 & 8 & 276 & All-to-All \\
 & 2021 & MI250X~\cite{MI250WhitePaper} & 128 & 3.2 & 16 & 800 & All-to-All \\
 & 2023 & MI300X~\cite{MI300XWhitePaper} & 192 & 5.3 & 256 & 896 & All-to-All \\
 & 2025 & MI355X~\cite{MI355XWhitePaper,amdMI355XDatasheet} & 288 & 8 & 256 & 1075.2 & All-to-All \\
 & 2026 & MI455X~\cite{amdMI455X2026,amdCDNA5} & 432 & 23.3 & 192 & 3600 & All-to-All \\
\midrule

\multirow{7}{*}{Google TPU}
 & 2021 & TPU~v4~\cite{2023ISCATPUv4} & 32 & 1.2 & -- & 300 & 3D mesh/torus \\
 & 2023 & TPU~v5e~\cite{tpuv5e} & 16 & 0.8 & -- & 400 & 2D torus \\
 & 2023 & TPU~v5p~\cite{tpuv5p} & 95 & 2.8 & -- & 1200 & 3D torus \\
 & 2024 & TPU~v6e~\cite{tpuv6e} & 32 & 1.6 & -- & 800 & 2D torus \\
 & 2025 & TPU~v7~\cite{tpuv7} & 192 & 7.4 & -- & 1200 & 3D torus \\
 & 2026 & TPU~8t~\cite{tpuv8} & 216 & 6.5 & 128 & 2400 & 3D torus \\
 & 2026 & TPU~8i~\cite{tpuv8} & 288 & 8.6 & 384 & 2400 & Boardfly \\
\midrule

\multirow{5}{*}{AWS Neuron}
 & 2018 & Inferentia 1~\cite{AWSInferentia, AWSInf1} & 8 & 0.05 & -- & 32 & 1D chain \\
 & 2020 & Trainium 1~\cite{AWSTrainuim, AWSTrn1} & 32 & 0.88 & -- & 384 & 2D torus \\
 & 2022 & Inferentia 2~\cite{AWSInferentia2, AWSInf2} & 32 & 0.88 & -- & 192 & 1D torus \\
 & 2023 & Trainium 2~\cite{AWSTrainuim2, AWSTrn2} & 96 & 2.9 & -- & 1280 & 2D torus\textsuperscript{b} \\
 & 2024 & Trainium 3~\cite{AWSTrainuim3, AWSTrn3} & 144 & 4.9 & -- & 2560 & All-to-All\textsuperscript{c} \\
\bottomrule
\end{tabular}%
}
\begin{minipage}{\textwidth}
\vspace{1mm}
\scriptsize
Topologies refer to representative platforms: P100/V100 to DGX-1 servers (V100 also appears in NVSwitch-based DGX-2~\cite{nvidiaHGX2Topology}); A100/H100 to DGX A100/H100 servers; B300/Rubin to GB300/Vera Rubin NVL72 racks. AMD MI50/MI100 refer to four-GPU ring/full-mesh hives; MI250X to four fully connected OAM modules; MI300X/MI355X to eight-GPU baseboards; MI455X to the 72-GPU Helios rack. TPU entries refer to supported pod slices. AWS entries refer to Inf1.24xlarge, Trn1.32xlarge, Inf2.48xlarge, Trn2 UltraServer, and Trn3 UltraServer, respectively~\cite{AWSInf1,AWSTrn1,AWSInf2,AWSTrn2,AWSTrn3}. Years follow first public family disclosure. GPU's on-chip SRAM refers to L2/shared cache, while NPU's on-chip SRAM refers to software-managed scratchpad.\par
\textsuperscript{a} Aggregate bidirectional bandwidth per accelerator as reported by the vendor. TPU 8 ICI uses 19.2 Tb/s $=2,400$ GB/s. Inferentia 1 uses DDR4; the other listed devices use HBM.\par
\textsuperscript{b} $4\times4$ 2D torus within a 16-chip Trn2 instance (1,024~GB/s per chip) plus same-coordinate rings across the four instances of a Trn2 UltraServer (256~GB/s per chip)~\cite{AWSTrn2}.\par
\textsuperscript{c} The interconnect bandwidth uses AWS's chip specification~\cite{AWSTrainuim3}.\par
\end{minipage}
\end{table*}




AI accelerators have strengthened their data-delivery capabilities to keep pace with rising computational throughput. As shown in Table~\ref{tab:mem_interconnect_vendor_multirow}, from the earliest to the latest listed NVIDIA and AMD GPU generations, off-chip high-bandwidth memory (HBM) capacity has increased by approximately 13.5--18$\times$, while HBM bandwidth has increased by approximately 23--27$\times$. HBM scaling is driven by two largely independent factors. Capacity improvements stem from (i) higher per-stack density, enabled by a greater number of dies per stack and increased DRAM die density, and (ii) a larger number of stacks integrated within a single package. Bandwidth gains arise from (i) higher per-pin signaling rates across successive HBM generations and (ii) expanded aggregate I/O width, primarily through additional stacks and, in some designs, wider interfaces. Chiplet-based packaging further supports these trends by enlarging the effective integration area and enhancing design flexibility, thereby facilitating the attachment of more memory stacks along with their associated PHYs and controllers.

\begin{figure}[ht]
    \centering
    \includegraphics[width=0.8\linewidth]{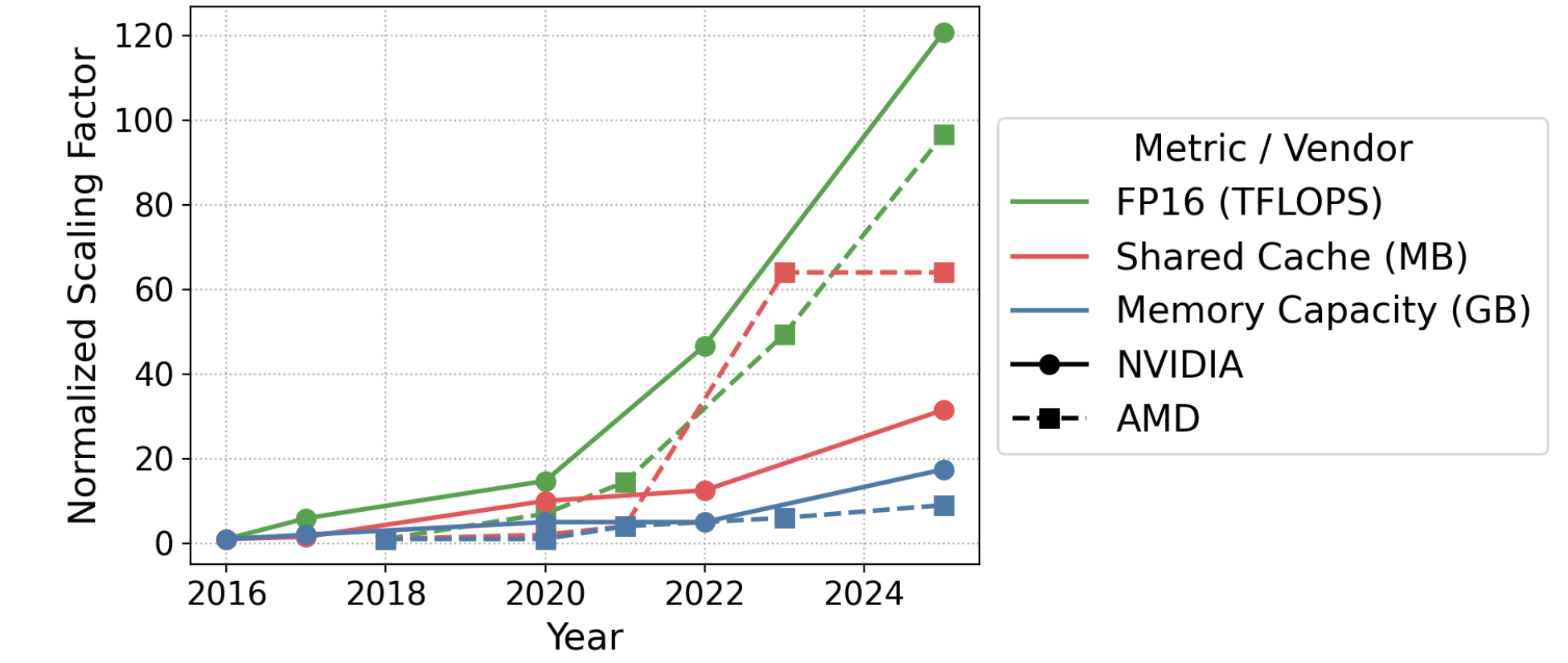}
    \caption{Scaling of per-GPU dense FP16/BF16 matrix throughput, shared-cache capacity, and HBM capacity across NVIDIA and AMD GPUs, 2016--2025. Each vendor/metric is normalized to its first observation.}
    \label{fig:FP16_memory}
    \Description{A line plot compares independently normalized FP16 throughput, shared-cache capacity, and HBM capacity from 2016 to 2025. NVIDIA uses solid lines and AMD dashed lines. The retained chart displays endpoint growth of approximately 120 and 96 times for FP16 throughput, 32 and 64 times for cache, and 18 and 9 times for HBM capacity, respectively; these plotted values predate the updated table specifications. Compute growth exceeds capacity growth for both vendors.}
\end{figure}

Because memory scaling has lagged behind computational growth, modern AI accelerators increasingly rely on larger on-chip SRAM to mitigate the widening performance gap. Through 2025, shared-cache capacity has increased by approximately 32$\times$ for NVIDIA GPUs and 64$\times$ for AMD GPUs relative to the earliest listed devices; here, ``shared cache'' refers to NVIDIA's L2 cache and AMD's large last-level cache. These expanded caches enhance data locality, amplify reuse, and alleviate pressure on off-chip memory bandwidth. Figure~\ref{fig:FP16_memory} compares the scaling trends of dense FP16/BF16 matrix throughput, shared-cache capacity, and HBM capacity for NVIDIA and AMD GPUs.

In the selected observations, total model parameters grow faster than per-accelerator memory capacity and the communication efficiency of scale-up fabrics (Figure~\ref{fig:scaling}). As a result, AI datacenters combine a scale-up fabric within each pod with a scale-out network across pods. Within the scale-up domain, interconnect bandwidth has increased steadily, while topologies have evolved to improve scalability and communication efficiency. AWS Trainium~\cite{AWSTrainuim, AWSTrainuim2, AWSTrainuim3} illustrates this trend. Trn1 and Trn2 connect 16 chips in a $4\times4$ 2D torus. Trn2 UltraServer links four such systems into a 64-chip domain through inter-instance rings. Trainium~3 replaces the torus with a switched all-to-all fabric, NeuronSwitch-v1, supporting up to 144 chips. The trade-off is straightforward. A low-radix torus avoids switch silicon and scales incrementally, but its diameter and contention increase with system size. A switched fabric requires more ports, silicon area, and power, but reduces path length and makes communication less sensitive to placement. This is particularly beneficial for communication-intensive collectives, such as mixture-of-experts (MoE) routing.

A switched all-to-all fabric is not the only alternative to a torus topology. Google TPU~8i~\cite{tpuv8} adopts Boardfly, a hierarchical high-radix topology built from four-chip building blocks. Eight boards form a fully connected group, and 36 such groups are interconnected through optical circuit switches, yielding 1,152 physical chip positions with support for up to 1,024 active chips. For the 1,024-chip configuration, Google reports a maximum path length of seven hops, compared with 16 hops for the referenced 3D torus, corresponding to a 56\% reduction. Shorter paths are particularly beneficial for latency-sensitive collectives: during autoregressive decoding, small reductions across model-parallel shards can make synchronization latency a substantial fraction of per-token execution time. Boardfly therefore reduces communication distance through hierarchical connectivity, whereas Trainium~3 UltraServers pursue a different approach based on a switched any-to-any fabric.

\subsection{Power Consumption and Cooling}

\label{sec:evolution_power_supply}

\begin{figure}[ht]
    \centering
    \includegraphics[width=\linewidth]{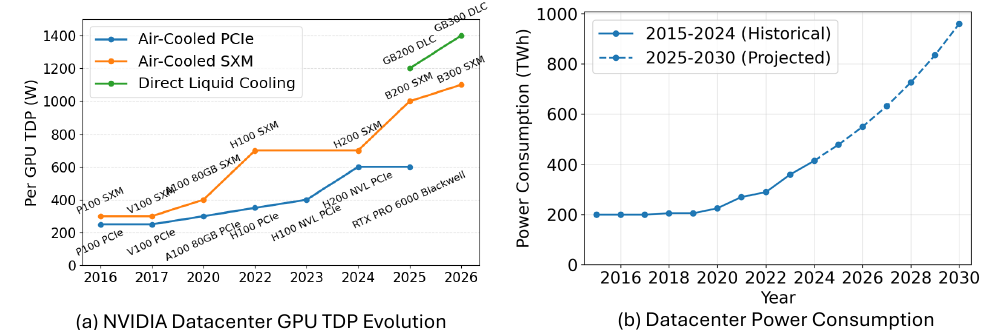}
    \caption{(a) Maximum reported NVIDIA accelerator power and cooling approach by product generation; power limits depend on the product and platform configuration. (b) Annual datacenter electricity consumption: the IEA supports the 2024 estimate of 415~TWh and 2030 projection of 945~TWh~\cite{iea_energy_ai_exec}; intervening projected values are an interpolation.}
    \label{fig:Power_increase}
    \Description{The left panel plots maximum reported NVIDIA GPU power against year: air-cooled PCIe devices rise from roughly 250 to 600 watts, air-cooled SXM devices from 300 to 1,100 watts, and the liquid-cooled observations reach 1,200 and 1,400 watts. GPU models are labeled at each observation. The right panel retains a solid historical energy series from about 200 TWh in 2015 to 415 TWh in 2024, followed by a dashed interpolation to the IEA 2030 projection of 945 TWh.}
\end{figure}

With the rapid advancement of AI accelerators, device-level power consumption has risen markedly. Over the past decade, the maximum power of high-end NVIDIA and AMD datacenter GPUs have increased from approximately 300~W to beyond 1000~W. This upward trend at the device level aggregates into substantial growth in datacenter electricity demand. The International Energy Agency (IEA) estimates that datacenters consumed roughly 415~TWh in 2024, accounting for about 1.5\% of global electricity use \cite{iea_energy_ai_exec,iea_energy_ai_demand}. Under this scenario, the IEA projects that the datacenter energy consumption will more than double to around 945~TWh by 2030, growing substantially faster than overall electricity demand.
Figure~\ref{fig:Power_increase}(b) summarizes historical trends and near-term projections of datacenter energy consumption.

Driven by the rapid escalation of accelerator power consumption, datacenter cooling has evolved through three distinct paradigms. Initially, the standard dual-slot, passively air-cooled PCIe form factor capped thermal design power (TDP) around the 300W mark for years, though recent iterations (e.g., H200 NVL) have stretched this to 600W. To break this thermal ceiling, the industry adopted planar-mounted modules such as NVIDIA's SXM and OCP's Open Accelerator Module (OAM), equipped with massive vertical tower heat sinks. This shift accommodated TDPs up to 1100W in recent generations like the B300 SXM. Today, the demand for larger parameter models necessitates even tighter cluster integration, accelerating the transition to direct liquid cooling (DLC). By eliminating bulky heatsinks, rack-scale DLC solutions (e.g., NVIDIA GB200) support extreme per-GPU TDPs up to 1400W while drastically shrinking the physical footprint. Crucially, this densification minimizes the copper wire length for scale-up networks like NVLink, thereby optimizing signal integrity and interconnect latency in massive AI clusters.

The rapid increase in datacenter electricity demand is also reflected in the scale of planned accelerator deployments. Public disclosures and credible reporting indicate that these deployments are reaching unprecedented levels: Meta planned to deploy approximately 350{,}000 NVIDIA H100 GPUs by the end of 2024, and about 600{,}000 H100-equivalent GPUs including other models \cite{reuters_meta_ai_chip_arsenal_2024}; xAI brought online a cluster comprising 100{,}000 NVIDIA Hopper GPUs \cite{nvidia_xai_colossus_2024}; and Oracle announced cloud superclusters scaling to 131{,}072 NVIDIA B200 GPUs \cite{oracle_supercluster_2024}. At such scale, infrastructure planning shifts from rack- or building-level considerations to campus-scale provisioning, requiring dedicated substations, expanded transmission capacity, and long-term power purchase agreements (PPAs), and in some cases on-site or near-site generation. As a result, power availability and grid interconnection timelines are increasingly becoming binding constraints on large-scale AI cluster deployment.

In response to these constraints, leading technology companies are coupling AI infrastructure expansion with long-horizon electricity procurement strategies and, in some cases, direct investment in generation assets. For example, Google announced significant AI-driven datacenter expansion alongside hydropower procurement at up to gigawatt scale to support rising load \cite{cnbc_google_pjm_2025,reuters_google_hydropower_2025}. Meta signed a 20-year agreement associated with the 1.1~GW Clinton Clean Energy Center, reflecting a shift toward firm, long-duration supply contracts structured around projected AI demand growth \cite{meta_constellation_2025,reuters_meta_nuclear_2025}. More broadly, hyperscalers including Amazon, Google, and Microsoft have pursued nuclear-linked procurement strategies, such as long-term offtake agreements and small modular reactor partnerships, to secure reliable low-carbon power for sustained AI expansion \cite{trellis_go_nuclear_2025}.

\section{Challenges of Future AI Datacenter Design}
\label{sec:future}

Designing future AI datacenters requires more than simply scaling today’s accelerators, networks, and racks. Every specialized architecture embodies assumptions about workload characteristics and their computational, memory, and communication requirements. Algorithms and hardware also co-evolve: emerging models and serving techniques alter these demands, while new hardware capabilities make previously impractical algorithms economically feasible. However, these components evolve on different timescales. Software and workloads may change during a deployment’s lifetime, whereas architecture decisions remain fixed for substantially longer periods.
The following subsections examine five related challenges: workload diversity; data placement and movement; communication across scale-up and scale-out networks; power delivery and cooling; and architectural specialization.

\subsection{Workload Specialization and Fleet Flexibility}
\label{sec:future_training_inference}
AI workload mix can change even when its installed accelerator capacity stays the same. Specializing hardware for particular workloads can improve efficiency, but can also make it harder to reassign capacity when demand shifts.

Training and inference each include workloads with different resource needs. Pre-training maintains optimizer state and uses large collective operations, while online inference is shaped by latency and throughput targets~\cite{duan2024training,li2024llminfer}. Inference is also varied and does not describe one fixed balance of hardware resources. Mixture-of-experts models, multimodal inputs, long contexts, retrieval augmentation, and diffusion place different demands on compute, memory capacity, memory bandwidth, and communication~\cite{ma2026inferencehardware}.
Current Transformer serving shows this variation within a single inference request. Prefill often offers more parallel work, while decode can spend more time moving weights and key--value (KV) cache. Draft and verify stages in speculative decoding, batching, prefix caching, prefill-decoding disaggregation, quantization, and context length further change this balance.

Google positions TPU~8t and TPU~8i for different workload mixes at the accelerator-pool level~\cite{tpuv8}, while OpenAI describes Jalape\~no as specialized for inference but balanced across a changing prefill/decode mix~\cite{openaiJalapeno2026}. These examples show that a system can be specialized for one broad class of work while retaining flexibility within that class.
Specialized resource pools introduce the challenge of matching available capacity to fluctuating demand. Maintaining separate pools for training and inference, or even for prefill and decode, can improve efficiency; however, one pool may remain underutilized while another experiences substantial queuing delays. 

\subsection{Data Placement and Movement}
\label{sec:future_data_movement}

AI workloads transfer large volumes of data across multiple memory tiers. For example, small-batch decoding can spend a substantial fraction of its execution time reading model weights and key-value (KV) caches~\cite{ma2026inferencehardware}. Placing data closer to compute units can reduce some of these transfers, but nearby memory capacity is limited, and data placement may require explicit software management. Architectures therefore shorten data-movement paths at different points in the memory hierarchy.

Groq and Cerebras provide on-chip SRAM, with Cerebras extending it across a wafer~\cite{Groq2020TSP,cerebras2023}. Retaining data in SRAM avoids repeated device-DRAM transfers. However, large parameter sizes, long contexts or greater request concurrency can exhaust the SRAM capacity. Offloading data to another memory tier introduces additional transfers, whereas distributing computation across more devices generates network communication. Under tensor parallelism, weight shards can remain on their assigned devices, but activations or partial results must still be exchanged. The memory organization ultimately determines which transfers remain necessary.

Processing-in-memory (PIM) reduces data movement by performing supported operations where the operands reside. d-Matrix Corsair computes within SRAM arrays~\cite{dMatrixcorsair}, whereas Samsung HBM-PIM and SK~hynix AiM integrate arithmetic units into DRAM devices~\cite{Samsungaquabolt,AiMJSSCC}. PIM architectures provide high internal memory bandwidth, but SRAM-based designs have limited capacity and DRAM-based designs have limited compute throughput.

Processing-near-memory (PNM) instead performs arithmetic in logic located close to memory, while 3D stacking can further shorten the links between the two. Raptor directly bonds logic to 3D DRAM to reduce the length of the memory interface~\cite{nair2026raptor}. Although this 3D-stacked architecture shortens data paths without compromising computational throughput, it also introduces thermal and reliability challenges.

\subsection{Communication Across Scale-Up and Scale-Out Networks}
\label{sec:future_communication}

Communication constraints arise from the interaction between workload mapping and physical network design. Large gradient exchanges in training can benefit from greater link bandwidth because transferring the payload occupies much of the communication time. Tensor-parallel decoding with small batches can instead exchange short messages repeatedly at each layer and token step~\cite{ma2026inferencehardware}. Faster links shorten payload transmission but do not remove the delay of starting an exchange, traversing the network, or waiting for other devices. The same fabric can therefore face different limits as the workload and batch size change.

Topology changes address part of this difference. Trainium moves from torus-based connectivity toward switched fabrics, adding switch hardware to reduce path lengths and contention~\cite{AWSTrainuim3}. Google reports that TPU~8i's Boardfly reduces the worst-case path across $\sim$1K chips from 16 hops in the compared 3D torus to 7~\cite{tpuv8}. This describes network diameter; the effect on a collective also depends on its schedule and placement. Scale-up fabrics provide high bandwidth and low latency within a server, rack, or pod, while scale-out networks connect these domains. If a model is partitioned across domains, exchanges between its partitions also use scale-out links. Expanding a scale-up domain can keep more exchanges on that fabric, but may require longer links. Copper connects nearby devices, while optical links extend reach. Co-packaged optics shorten electrical paths between switches and optical interfaces~\cite{nvidiaCPO2025}; they reduce part of the signaling cost without removing propagation delay or workload synchronization.

\subsection{Power Delivery and Cooling from Rack to Facility}
\label{sec:future_cooling}

Power delivery and cooling can limit how much installed compute capacity a datacenter can use. Sufficient total power does not guarantee that each rack can receive that power or dissipate the resulting heat.

At a fixed distribution voltage, increasing rack power raises the current carried by the conductors. For a 200~kW rack supplied at 54~V, $I=P/V$ yields approximately 3,700~A. Increasing current through a conductor raises resistive losses and heat generation, whereas increasing the conductor cross-section requires additional material and space. NVIDIA identifies physical limitations of 54~VDC distribution as rack power exceeds 200~kW and proposes an 800~VDC Architecture~\cite{nvidia800VDC2025}. A higher distribution voltage reduces the current required to deliver a given amount of power. Consequently, upgrading accelerators may require corresponding changes to rack- and facility-level power-delivery infrastructure even when sufficient floor space is available.

Cooling similarly links package design to facility-level energy consumption. NVIDIA specifies a maximum inlet-water temperature of 45$^\circ$C for its Vera Rubin MGX rack design, potentially reducing chiller use in climates where outdoor air can dissipate much of the heat~\cite{nvidiaCoolingEfficiency2026}. For a fixed thermal resistance between the chip and coolant, however, warmer water reduces the temperature margin below the chip’s operating limit. As the device approaches this limit, sustaining higher activity requires either improved heat transfer or lower coolant temperatures. Thus, the available electrical-power budget alone does not determine how much accelerator activity a cooling system can sustain.

Power demand also fluctuates over the course of a job. Synchronized training workloads alternate between compute-intensive and communication-intensive phases, producing coordinated power swings across many GPUs~\cite{choukse2025power}. Short-term energy storage can buffer these fluctuations but has finite capacity, whereas throttling or delaying computation can reduce peak demand at the cost of slower job progress. Consequently, two workloads with similar average power consumption may impose substantially different demands on electrical infrastructure depending on the magnitude, timing, and synchronization of their power peaks.

\subsection{Architectural Specialization and Model Evolution}
\label{sec:future_fixed_parameters}

Architectural specialization differs in how much of a model's computation is fixed in hardware and how much remains programmable. This affects which model updates can be handled in software and which require changes to the chip.
In SambaNova’s SN40L, the compiler maps supported model graphs onto configurable compute and memory tiles and programs the corresponding data routes~\cite{MICRO2024sambanova}. Reconfigurability thus provides flexibility for algorithmic evolution, but only within a finite set of physical resources and supported operations.

Taalas HC1 adopts a different approach by hard-wiring the AI model while retaining configurable context lengths and low-rank adapters for fine-tuning~\cite{taalas2026ubiquitous}. Updating an adapter can modify model behavior, whereas replacing the hard-wired base model requires new silicon. Existing chips can continue serving the original model as long as demand persists, but their capacity becomes difficult to repurpose if demand shifts to an unsupported model. Encoding model-specific functionality in hardware can simplify execution, but it also makes hardware reuse dependent on which aspects remain programmable. Across these examples, software can reassign work, remap computation, or constrain activity, but the cost and effectiveness of such adaptations are determined by the physical resources already deployed.

\section{Conclusion}
\label{sec:conclusion}

This survey has examined industrial AI datacenter architectures through accelerator organization, scale-up interconnects, and architectural evolution. Across the surveyed systems, specialized arithmetic coexists with general-purpose execution resources and varied memory-management strategies. The interconnect analysis relates collective communication to node-, rack-, and pod-scale fabrics, while the generational studies connect gains in computational throughput with changes in operand delivery, memory hierarchy, communication, and cooling.

These comparisons show that specialization cannot be assessed from the compute engine alone. Reducing memory traffic can increase sensitivity to data placement, while distributing models across devices introduces communication overheads that faster arithmetic cannot eliminate. Power and thermal constraints further limit usable system capacity. As workloads evolve, these architectural choices also affect how effectively deployed hardware can be reused, linking efficiency with system flexibility and longevity.

\bibliographystyle{ACM-Reference-Format}
\bibliography{sample-base}

@online{ocpMX2023,
  author       = {{Open Compute Project}},
  title        = {{OCP Microscaling Formats (MX) Specification}},
  year         = {2023},
  month        = sep,
  note         = {Version 1.0},
  url          = {https://www.opencompute.org/documents/ocp-microscaling-formats-mx-v1-0-spec-final-pdf},
  urldate      = {2026-09-13}
}

@online{nvidiaNVFP4Format2025,
  author       = {Alvarez, Eduardo and Almog, Omri and Chung, Eric and Layton, Simon and Stosic, Dusan and Krashinsky, Ronny and Aubrey, Kyle},
  title        = {{Introducing {NVFP4} for Efficient and Accurate Low-Precision Inference}},
  year         = {2025},
  month        = jun,
  url          = {https://developer.nvidia.com/blog/introducing-nvfp4-for-efficient-and-accurate-low-precision-inference/},
  urldate      = {2026-09-13}
}

@online{nvidiaHGX2Topology,
  author       = {Tsu, William},
  title        = {{{HGX-2} Fuses {HPC} and {AI} Computing Architectures}},
  year         = {2018},
  month        = may,
  url          = {https://developer.nvidia.com/blog/hgx-2-fuses-ai-computing/},
  urldate      = {2026-09-13}
}

@online{tenstorrentGalaxyMesh,
  author       = {{Tenstorrent}},
  title        = {{Programming Multiple Meshes}},
  url          = {https://github.com/tenstorrent/tt-metal/blob/main/tech_reports/Programming_Multiple_Meshes/Programming_Multiple_Meshes.md},
  urldate      = {2026-09-20},
  note = {Undated documentation; no publication year supplied}
}

@misc{dighe2026maia200,
  author       = {Dighe, Saurabh and Levin, Artour},
  title        = {{Deep Dive into the Maia 200 Architecture}},
  howpublished = {\url{https://techcommunity.microsoft.com/blog/azureinfrastructureblog/deep-dive-into-the-maia-200-architecture/4489312}},
  institution  = {Microsoft},
  note         = {Azure Infrastructure Blog. Published Jan. 26, 2026; updated Jan. 30, 2026},
  year         = {2026},
  month        = jan,
  urldate      = {2026-06-29}
}

@online{cambriconBangCGuide,
  author       = {Cambricon},
  title        = {{BANG C Language Developer Guide}},
  year         = {2020},
  note         = {Version 2.4.1, February 21, 2020},
  url          = {https://forum.cambricon.com/uploadfile/user/file/20201125/1606289569710855.pdf},
}

@online{NVIDIA-groq-3,
  author       = {Aubrey, Kyle and Ghodsian, Farshad},
  title        = {{Inside {NVIDIA Groq 3 LPX}: The Low-Latency Inference Accelerator for the {NVIDIA Vera Rubin} Platform}},
  year         = {2026},
  month        = mar,
  url          = {https://developer.nvidia.com/blog/inside-nvidia-groq-3-lpx-the-low-latency-inference-accelerator-for-the-nvidia-vera-rubin-platform/},
}

@online{HopperTMA,
  author       = {Hoque, Adnan and Wright, Less and Yang, Chih-Chieh},
  title        = {{Deep Dive on the Hopper TMA Unit for FP8 GEMMs}},
  year         = {2024},
  month        = jul,
  url          = {https://pytorch.org/blog/hopper-tma-unit/}, 
}

@online{AWSInferentia,
  author       = {AWS},
  title        = {{Inferentia Architecture}},
  year         = {2026},
  note         = {AWS Neuron SDK 2.31.1 documentation},
  url          = {https://awsdocs-neuron.readthedocs-hosted.com/en/v2.31.1/about-neuron/arch/neuron-hardware/inferentia.html},
}

@online{AWSInferentia2,
  author       = {AWS},
  title        = {{Inferentia2 Architecture}},
  year         = {2026},
  note         = {AWS Neuron SDK 2.31.1 documentation},
  url          = {https://awsdocs-neuron.readthedocs-hosted.com/en/v2.31.1/about-neuron/arch/neuron-hardware/inferentia2.html},
}

@online{AWSTrainuim,
  author       = {AWS},
  title        = {{Trainium Architecture}},
  year         = {2026},
  note         = {AWS Neuron SDK 2.31.1 documentation},
  url          = {https://awsdocs-neuron.readthedocs-hosted.com/en/v2.31.1/about-neuron/arch/neuron-hardware/trainium.html},
}

@online{AWSTrainuim2,
  author       = {AWS},
  title        = {{Trainium2 Architecture}},
  year         = {2026},
  note         = {AWS Neuron SDK 2.31.1 documentation},
  url          = {https://awsdocs-neuron.readthedocs-hosted.com/en/v2.31.1/about-neuron/arch/neuron-hardware/trainium2.html},
}

@online{AWSTrainuim3,
  author       = {AWS},
  title        = {{Trainium3 Architecture}},
  year         = {2026},
  note         = {AWS Neuron SDK 2.31.1 documentation},
  url          = {https://awsdocs-neuron.readthedocs-hosted.com/en/v2.31.1/about-neuron/arch/neuron-hardware/trainium3.html},
}

@online{NeuronCorev2,
  author       = {AWS},
  title        = {{NeuronCore-v2 Architecture}},
  year         = {2026},
  note         = {AWS Neuron SDK 2.31.1 documentation},
  url          = {https://awsdocs-neuron.readthedocs-hosted.com/en/v2.31.1/about-neuron/arch/neuron-hardware/neuron-core-v2.html},
}

@online{NeuronCorev4,
  author       = {AWS},
  title        = {{NeuronCore-v4 Architecture}},
  year         = {2026},
  note         = {AWS Neuron SDK 2.31.1 documentation},
  url          = {https://awsdocs-neuron.readthedocs-hosted.com/en/v2.31.1/about-neuron/arch/neuron-hardware/neuron-core-v4.html},
}

@online{AWSInf1,
  author       = {AWS},
  title        = {{Amazon EC2 Inf1 Architecture}},
  year         = {2026},
  note         = {AWS Neuron SDK 2.31.1 documentation},
  url          = {https://awsdocs-neuron.readthedocs-hosted.com/en/v2.31.1/about-neuron/arch/neuron-hardware/inf1-arch.html},
}

@online{AWSInf2,
  author       = {AWS},
  title        = {{Amazon EC2 Inf2 Architecture}},
  year         = {2026},
  note         = {AWS Neuron SDK 2.31.1 documentation},
  url          = {https://awsdocs-neuron.readthedocs-hosted.com/en/v2.31.1/about-neuron/arch/neuron-hardware/inf2-arch.html},
}

@online{AWSTrn1,
  author       = {AWS},
  title        = {{Amazon EC2 Trn1 Architecture}},
  year         = {2026},
  note         = {AWS Neuron SDK 2.31.1 documentation},
  url          = {https://awsdocs-neuron.readthedocs-hosted.com/en/v2.31.1/about-neuron/arch/neuron-hardware/trn1-arch.html},
}

@online{AWSTrn2,
  author       = {AWS},
  title        = {{Amazon EC2 Trn2 Architecture}},
  year         = {2026},
  note         = {AWS Neuron SDK 2.31.1 documentation},
  url          = {https://awsdocs-neuron.readthedocs-hosted.com/en/v2.31.1/about-neuron/arch/neuron-hardware/trn2-arch.html},
}

@online{AWSTrn3,
  author       = {AWS},
  title        = {{Amazon EC2 Trn3 Architecture}},
  year         = {2026},
  note         = {AWS Neuron SDK 2.31.1 documentation},
  url          = {https://awsdocs-neuron.readthedocs-hosted.com/en/v2.31.1/about-neuron/arch/neuron-hardware/trn3-arch.html},
}

@online{tpuArch,
  author       = {{Google Cloud}},
  title        = {{TPU architecture}},
  year         = {2026},
  url          = {https://docs.cloud.google.com/tpu/docs/system-architecture-tpu-vm},
}

@online{tpuv5e,
  author       = {{Google Cloud}},
  title        = {{TPU v5e}},
  year         = {2026},
  url          = {https://docs.cloud.google.com/tpu/docs/v5e}, 
}

@online{tpuv5p,
  author       = {{Google Cloud}},
  title        = {{TPU v5p}},
  year         = {2026},
  url          = {https://docs.cloud.google.com/tpu/docs/v5p}, 
}

@online{tpuv6e,
  author       = {{Google Cloud}},
  title        = {{TPU v6e}},
  year         = {2026},
  url          = {https://docs.cloud.google.com/tpu/docs/v6e}, 
}

@online{tpuv7,
  author       = {{Google Cloud}},
  title        = {{TPU7x (Ironwood)}},
  year         = {2026},
  url          = {https://docs.cloud.google.com/tpu/docs/tpu7x}, 
}

@online{tpuv8,
  author       = {{Google Cloud}},
  title        = {{Inside the eighth-generation TPU: An architecture deep dive}},
  year         = {2026},
  month        = apr,
  url          = {https://cloud.google.com/blog/products/compute/tpu-8t-and-tpu-8i-technical-deep-dive}, 
}

@online{MI100WhitePaper,
  author       = {AMD},
  title        = {{Introducing AMD CDNA Architecture: The All-New AMD GPU Architecture for the Modern Era of HPC \& AI }},
  year         = {2020},
  url          = {https://www.amd.com/content/dam/amd/en/documents/instinct-business-docs/white-papers/amd-cdna-white-paper.pdf}, 
}

@online{MI250WhitePaper,
  author       = {AMD},
  title        = {{Introducing AMD CDNA 2 Architecture: Propelling humanity’s foremost research with the world’s most powerful HPC and AI accelerator.}},
  year         = {2021},
  url          = {https://www.amd.com/content/dam/amd/en/documents/instinct-business-docs/white-papers/amd-cdna2-white-paper.pdf}, 
}

@online{MI300XWhitePaper,
  author       = {AMD},
  title        = {{Introducing AMD CDNA 3 Architecture: The All-New AMD GPU Architecture for the Modern Era of HPC \& AI}},
  year         = {2025},
  url          = {https://www.amd.com/content/dam/amd/en/documents/instinct-tech-docs/white-papers/amd-cdna-3-white-paper.pdf}, 
}

@online{MI355XWhitePaper,
  author       = {AMD},
  title        = {{Introducing AMD CDNA 4 Architecture: Breakthrough AI and HPC Acceleration with Enhanced AI
Capabilities, Advanced Precisions, and High Efficiency}},
  year         = {2025},
  url          = {https://www.amd.com/content/dam/amd/en/documents/instinct-tech-docs/white-papers/amd-cdna-4-architecture-whitepaper.pdf}, 
}

@online{A100WhitePaper,
  author       = {NVIDIA},
  title        = {{NVIDIA A100 Tensor Core GPU Architecture}},
  year         = {2020},
  url          = {https://images.nvidia.com/aem-dam/en-zz/Solutions/data-center/nvidia-ampere-architecture-whitepaper.pdf}, 
}

@online{CuTeMMA,
  author       = {NVIDIA},
  title        = {{CuTe’s support for Matrix Multiply-Accumulate instructions}},
  year         = {2026},
  url          = {https://docs.nvidia.com/cutlass/latest/media/docs/cpp/cute/0t_mma_atom.html}, 
}

@online{B300WhitePaper,
  author       = {NVIDIA},
  title        = {{NVIDIA Blackwell Ultra}},
  year         = {2025},
  url          = {https://resources.nvidia.com/en-us-blackwell-architecture/blackwell-ultra-datasheet},
  note = {October 2025 datasheet, GB300 NVL72 configuration; dense throughput follows the datasheet footnotes},
  urldate = {2026-09-20}
}

@online{blackwell,
  author       = {Aubrey, Kyle and Stam, Nick},
  title        = {{Inside NVIDIA Blackwell Ultra: The Chip Powering the AI Factory Era}},
  year         = {2025},
  month        = aug,
  url          = {https://developer.nvidia.com/blog/inside-nvidia-blackwell-ultra-the-chip-powering-the-ai-factory-era/}, 
}

@online{nvshmem,
  author       = {NVIDIA},
  title        = {{Programming Model Overview}},
  year         = {2020},
  note         = {NVSHMEM 1.0.1 documentation},
  url          = {https://docs.nvidia.com/nvshmem/archives/nvshmem-101/api/docs/gen/overview.html}, 
}

@online{nvswitch,
  author       = {NVIDIA},
  title        = {{NVIDIA NVSWITCH: The World’s Highest-Bandwidth On-Node Switch}},
  year         = {2018},
  url          = {https://images.nvidia.com/content/pdf/nvswitch-technical-overview.pdf}, 
}

@online{sparsity,
  author       = {Pool, Jeff and Sawarkar, Abhishek and Rodge, Jay},
  title        = {{Accelerating Inference with Sparsity Using the NVIDIA Ampere Architecture and NVIDIA TensorRT}},
  year         = {2021},
  month        = jul,
  url          = {https://developer.nvidia.com/blog/accelerating-inference-with-sparsity-using-ampere-and-tensorrt/},
}

@article{nvidiaA100Paper,
  title={{NVIDIA A100 tensor core GPU: Performance and innovation}},
  author={Choquette, Jack and Gandhi, Wishwesh and Giroux, Olivier and Stam, Nick and Krashinsky, Ronny},
  journal={IEEE Micro},
  volume={41},
  number={2},
  pages={29--35},
  year={2021},
  publisher={IEEE}
}

@article{nvidiaH100Paper,
  title={{NVIDIA hopper H100 GPU: Scaling performance}},
  author={Choquette, Jack},
  journal={IEEE Micro},
  volume={43},
  number={3},
  pages={9--17},
  year={2023},
  publisher={IEEE}
}

@article{nvidiaV100Paper,
  title={{Volta: Performance and programmability}},
  author={Choquette, Jack and Giroux, Olivier and Foley, Denis},
  journal={IEEE Micro},
  volume={38},
  number={2},
  pages={42--52},
  year={2018},
  publisher={IEEE}
}

@article{nvidiaP100Paper,
  title={{Ultra-performance Pascal GPU and NVLink interconnect}},
  author={Foley, Denis and Danskin, John},
  journal={IEEE Micro},
  volume={37},
  number={2},
  pages={7--17},
  year={2017},
  publisher={IEEE}
}

@article{amdMI300XPaper,
  title={{AMD Instinct™ MI300X: A Generative AI Accelerator and Platform Architecture}},
  author={Smith, Alan and Alla, Vamsi Krishna},
  journal={IEEE Micro},
  year={2025},
  publisher={IEEE},
  volume = {45},
  number = {3},
  pages = {41--48},
  doi = {10.1109/mm.2025.3552324}
}

@inproceedings{2023ISCATPUv4,
  title={{TPU v4: An optically reconfigurable supercomputer for machine learning with hardware support for embeddings}},
  author={Jouppi, Norm and Kurian, George and Li, Sheng and Ma, Peter and Nagarajan, Rahul and Nai, Lifeng and Patil, Nishant and Subramanian, Suvinay and Swing, Andy and Towles, Brian and others},
  booktitle={Proceedings of the 50th annual international symposium on computer architecture},
  pages={1--14},
  year={2023}
}

@article{TPUv2v3,
  title={{The design process for Google's training chips: TPU v2 and TPU v3}},
  author={Norrie, Thomas and Patil, Nishant and Yoon, Doe Hyun and Kurian, George and Li, Sheng and Laudon, James and Young, Cliff and Jouppi, Norman and Patterson, David},
  journal={IEEE Micro},
  volume={41},
  number={2},
  pages={56--63},
  year={2021},
  publisher={IEEE}
}

@inproceedings{IntelGaudi3HC36,
  title={{Intel Gaudi 3 AI Accelerator: Architected for Gen AI Training and Inference}},
  author={Kaplan, Roman},
  booktitle={2024 IEEE Hot Chips 36 Symposium (HCS)},
  pages={1--16},
  year={2024},
  organization={IEEE},
  publisher = {IEEE}
}

@article{TeslaDojo,
  title={{The Microarchitecture of Dojo, Tesla's Exa-Scale Computer}},
  author={Talpes, Emil and Sarma, Debjit Das and Williams, Doug and Arora, Sahil and Kunjan, Thomas and Floering, Benjamin and Jalote, Ankit and Hsiong, Christopher and Poorna, Chandrasekhar and Samant, Vaidehi and others},
  journal={IEEE Micro},
  volume={43},
  number={3},
  pages={31--39},
  year={2023},
  publisher={IEEE}
}

@inproceedings{MicrosoftMaia,
  title={{Inside Maia 100}},
  author={Xu, Sherry and Ramakrishnan, Chandru},
  booktitle={2024 IEEE Hot Chips 36 Symposium (HCS)},
  pages={1--17},
  year={2024},
  organization={IEEE Computer Society},
  publisher = {IEEE}
}

@inproceedings{MICRO2024sambanova,
  title={{SambaNova SN40L: Scaling the AI Memory Wall with Dataflow and Composition of Experts}},
  author={Prabhakar, Raghu and Sivaramakrishnan, Ram and Gandhi, Darshan and Du, Yun and Wang, Mingran and Song, Xiangyu and Zhang, Kejie and Gao, Tianren and Wang, Angela and Li, Xiaoyan and others},
  booktitle={2024 57th IEEE/ACM International Symposium on Microarchitecture (MICRO)},
  pages={1353--1366},
  year={2024},
  organization={IEEE},
  publisher = {IEEE}
}

@article{HuaweiAscend,
  title={{AIBench: A Tool for Benchmarking Huawei Ascend AI Processors}},
  author={Xiao, Yang and Wang, Zeke},
  journal={CCF Transactions on High Performance Computing},
  volume={6},
  number={2},
  pages={115--129},
  year={2024},
  publisher={Springer}
}

@inproceedings{vasiljevic2024blackhole,
  title={{Blackhole \& TT-Metalium: The Standalone AI Computer and Its Programming Model}},
  author={Vasiljevic, Jasmina and Capalija, Davor},
  booktitle={2024 IEEE Hot Chips 36 Symposium (HCS)},
  pages={1--30},
  year={2024},
  organization={IEEE Computer Society Los Alamitos, CA, USA},
  publisher = {IEEE}
}

@inproceedings{MTIA,
  title={{MTIA: First Generation Silicon Targeting Meta's Recommendation Systems}},
  author={Firoozshahian, Amin and Coburn, Joel and Levenstein, Roman and Nattoji, Rakesh and Kamath, Ashwin and Wu, Olivia and Grewal, Gurdeepak and Aepala, Harish and Jakka, Bhasker and Dreyer, Bob and others},
  booktitle={Proceedings of the 50th Annual International Symposium on Computer Architecture},
  pages={1--13},
  year={2023}
}

@inproceedings{MTIA2,
  title={{Meta's Second Generation AI Chip: Model-Chip Co-Design and Productionization Experiences}},
  author={Coburn, Joel and Tang, Chunqiang and Asal, Sameer Abu and Agrawal, Neeraj and Chinta, Raviteja and Dixit, Harish and Dodds, Brian and Dwarakapuram, Saritha and Firoozshahian, Amin and Gao, Cao and others},
  booktitle={Proceedings of the 52nd Annual International Symposium on Computer Architecture},
  pages={1689--1702},
  year={2025}
}

@inproceedings{qualcommAI100,
  title={{Qualcomm{\textregistered} Cloud AI 100: 12TOPS/W Scalable, High Performance and Low Latency Deep Learning Inference Accelerator}},
  author={Chatha, Karam},
  booktitle={2021 IEEE Hot Chips 33 Symposium (HCS)},
  pages={1--19},
  year={2021},
  organization={IEEE},
  publisher = {IEEE}
}

@inproceedings{Groq2022software,
  title={{A software-defined tensor streaming multiprocessor for large-scale machine learning}},
  author={Abts, Dennis and Kimmell, Garrin and Ling, Andrew and Kim, John and Boyd, Matt and Bitar, Andrew and Parmar, Sahil and Ahmed, Ibrahim and DiCecco, Roberto and Han, David and others},
  booktitle={Proceedings of the 49th Annual International Symposium on Computer Architecture},
  pages={567--580},
  year={2022},
  doi={10.1145/3470496.3527405}
}

@inproceedings{Groq2020TSP,
  title={{Think fast: A tensor streaming processor (TSP) for accelerating deep learning workloads}},
  author={Abts, Dennis and Ross, Jonathan and Sparling, Jonathan and Wong-VanHaren, Mark and Baker, Max and Hawkins, Tom and Bell, Andrew and Thompson, John and Kahsai, Temesghen and Kimmell, Garrin and others},
  booktitle={2020 ACM/IEEE 47th Annual International Symposium on Computer Architecture (ISCA)},
  pages={145--158},
  year={2020},
  organization={IEEE},
  publisher = {IEEE}
}

@article{cerebras2023,
  title={{Cerebras architecture deep dive: First look inside the hardware/software co-design for deep learning}},
  author={Lie, Sean},
  journal={IEEE Micro},
  volume={43},
  number={3},
  pages={18--30},
  year={2023},
  publisher={IEEE}
}

@misc{dissectingGraphcore,
  title={{Dissecting the graphcore ipu architecture via microbenchmarking}},
  author={Jia, Zhe and Tillman, Blake and Maggioni, Marco and Scarpazza, Daniele Paolo},


  year={2019},
  archivePrefix = {arXiv},
  eprint = {1912.03413},
  url = {https://arxiv.org/abs/1912.03413}
}

@article{dMatrixcorsair,
  title={{Corsair: An In-memory Computing Chiplet Architecture for Inference-time Compute Acceleration}},
  author={Srivastava, Satyam and Arunkumar, Akhil and Kurella, Nithesh and Panda, Amrit and Jain, Gaurav and Kamath, Purushotham and Wutzke, Mark and Tiruvur, Arun and Gupta, Mike and Soloveychik, Ilya and others},
  journal={IEEE Micro},
  volume={45},
  number={5},
  pages={30--42},
  year={2025},
  publisher={IEEE},
  doi={10.1109/MM.2025.3593444}
}

@inproceedings{AimISSCC,
  title={{A 1ynm 1.25 V 8Gb, 16Gb/s/pin GDDR6-based accelerator-in-memory supporting 1TFLOPS MAC operation and various activation functions for deep-learning applications}},
  author={Lee, Seongju and Kim, Kyuyoung and Oh, Sanghoon and Park, Joonhong and Hong, Gimoon and Ka, Dongyoon and Hwang, Kyudong and Park, Jeongje and Kang, Kyeongpil and Kim, Jungyeon and others},
  booktitle={2022 IEEE International Solid-State Circuits Conference (ISSCC)},
  volume={65},
  pages={1--3},
  year={2022},
  organization={IEEE},
  doi={10.1109/ISSCC42614.2022.9731711},
  publisher = {IEEE}
}

@article{AiMJSSCC,
  title={{A 1ynm 1.25 v 8Gb 16Gb/s/pin GDDR6-based accelerator-in-memory supporting 1TFLOPS MAC operation and various activation functions for deep learning application}},
  author={Kwon, Daehan and Lee, Seongju and Kim, Kyuyoung and Oh, Sanghoon and Park, Joonhong and Hong, Gi-Moon and Ka, Dongyoon and Hwang, Kyudong and Park, Jeongje and Kang, Kyeongpil and others},
  journal={IEEE Journal of Solid-State Circuits},
  volume={58},
  number={1},
  pages={291--302},
  year={2023},
  publisher={IEEE},
  doi={10.1109/JSSC.2022.3200718}
}

@article{Samsungaquabolt,
  title={{Aquabolt-XL HBM2-PIM, LPDDR5-PIM with in-memory processing, and AXDIMM with acceleration buffer}},
  author={Kim, Jin Hyun and Kang, Shin-Haeng and Lee, Sukhan and Kim, Hyeonsu and Ro, Yuhwan and Lee, Seungwon and Wang, David and Choi, Jihyun and So, Jinin and Cho, YeonGon and others},
  journal={IEEE Micro},
  volume={42},
  number={3},
  pages={20--30},
  year={2022},
  publisher={IEEE},
  doi={10.1109/MM.2022.3164651}
}

@inproceedings{krizhevsky2012alexnet,
  title     = {{ImageNet Classification with Deep Convolutional Neural Networks}},
  author    = {Krizhevsky, Alex and Sutskever, Ilya and Hinton, Geoffrey E.},
  booktitle = {Advances in Neural Information Processing Systems},
  year      = {2012},
  volume = {25},
  publisher = {Curran Associates, Inc.},
  numpages = {9},
  url = {https://proceedings.neurips.cc/paper_files/paper/2012/file/c399862d3b9d6b76c8436e924a68c45b-Paper.pdf}
}

@article{sun2022dissecting,
  title={{Dissecting tensor cores via microbenchmarks: Latency, throughput and numeric behaviors}},
  author={Sun, Wei and Li, Ang and Geng, Tong and Stuijk, Sander and Corporaal, Henk},
  journal={IEEE Transactions on Parallel and Distributed Systems},
  volume={34},
  number={1},
  pages={246--261},
  year={2023},
  publisher={IEEE},
  doi = {10.1109/TPDS.2022.3217824},
  url = {https://arxiv.org/abs/2206.02874v3},
  note = {Appendix A: \url{https://arxiv.org/abs/2206.02874v3}}
}

@inproceedings{luo2024benchmarking,
  title={{Benchmarking and Dissecting the NVIDIA Hopper GPU Architecture}},
  author={Luo, Weile and Fan, Ruibo and Li, Zeyu and Du, Dayou and Wang, Qiang and Chu, Xiaowen},
  booktitle={2024 IEEE International Parallel and Distributed Processing Symposium (IPDPS)},
  pages={656--667},
  year={2024},
  organization={IEEE},
  publisher = {IEEE}
}

@misc{jarmusch2025microbenchmarking,
  title={{Microbenchmarking NVIDIA's Blackwell Architecture: An in-depth Architectural Analysis}},
  author={Jarmusch, Aaron and Chandrasekaran, Sunita},
  year={2025},
  eprint = {2512.02189v1},
  archivePrefix = {arXiv},
  url = {https://arxiv.org/abs/2512.02189v1},
  note = {Version 1, December 1, 2025}
}

@misc{sparseDNN,
  title={{Accelerating sparse deep neural networks}},
  author={Mishra, Asit and Latorre, Jorge Albericio and Pool, Jeff and Stosic, Darko and Stosic, Dusan and Venkatesh, Ganesh and Yu, Chong and Micikevicius, Paulius},


  year={2021},
  archivePrefix = {arXiv},
  eprint = {2104.08378},
  url = {https://arxiv.org/abs/2104.08378}
}

@inproceedings{yan2020demystifying,
  title={{Demystifying tensor cores to optimize half-precision matrix multiply}},
  author={Yan, Da and Wang, Wei and Chu, Xiaowen},
  booktitle={2020 IEEE International Parallel and Distributed Processing Symposium (IPDPS)},
  pages={634--643},
  year={2020},
  organization={IEEE},
  publisher = {IEEE}
}

@online{iea_energy_ai_exec,
  author  = {{International Energy Agency}},
  title   = {{Energy and AI: Executive Summary}},
  year    = {2025},
  url     = {https://www.iea.org/reports/energy-and-ai/executive-summary},
  urldate = {2026-02-17}
}

@online{iea_energy_ai_demand,
  author  = {{International Energy Agency}},
  title   = {{Energy and AI: Energy demand from AI}},
  year    = {2025},
  url     = {https://www.iea.org/reports/energy-and-ai/energy-demand-from-ai},
  urldate = {2026-02-17}
}

@online{cnbc_google_pjm_2025,
  author  = {{CNBC}},
  title   = {{Google to invest \$25 billion in data centers, AI infrastructure in PJM}},
  year    = {2025},
  month   = jul,
  url     = {https://www.cnbc.com/2025/07/15/google-to-invest-25-billion-in-data-centers-ai-infrastructure-in-pjm.html},
  urldate = {2026-02-17}
}

@online{trellis_go_nuclear_2025,
  author  = {Clancy, Heather},
  title   = {{Amazon, Google, Meta and Microsoft go nuclear}},
  year    = {2025},
  month   = jun,
  url     = {https://trellis.net/article/amazon-google-meta-and-microsoft-go-nuclear/},
  urldate = {2026-02-17}
}

@online{reuters_google_hydropower_2025,
  author  = {{Reuters}},
  title   = {{Google inks \$3 billion US hydropower deal in largest clean energy agreement of its kind}},
  year    = {2025},
  month   = jul,
  url     = {https://www.reuters.com/sustainability/boards-policy-regulation/google-inks-3-billion-us-hydropower-deal-largest-clean-energy-agreement-its-kind-2025-07-15/},
  urldate = {2026-02-17}
}

@online{reuters_meta_nuclear_2025,
  author  = {{Reuters}},
  title   = {{Meta signs power agreement with Constellation nuclear plant}},
  year    = {2025},
  month   = jun,
  url     = {https://www.reuters.com/sustainability/climate-energy/meta-signs-power-agreement-with-constellation-nuclear-plant-2025-06-03/},
  urldate = {2026-02-17}
}

@online{meta_constellation_2025,
  author  = {{Constellation Energy}},
  title   = {{Constellation, Meta Sign 20-Year Deal for Clean, Reliable Nuclear Energy in Illinois}},
  year    = {2025},
  month   = jun,
  url     = {https://www.constellationenergy.com/newsroom/2025/constellation-meta-sign-20-year-deal-for-clean-reliable-nuclear-energy-in-illinois.html},
  urldate = {2026-02-17}
}

@online{reuters_meta_ai_chip_arsenal_2024,
  author  = {{Reuters}},
  title   = {{Meta ramps up AI efforts by building chip arsenal}},
  year    = {2024},
  month   = jan,
  url     = {https://www.reuters.com/technology/meta-bringing-together-ai-research-product-teams-2024-01-18/},
  urldate = {2026-02-17}
}

@online{nvidia_xai_colossus_2024,
  author  = {{NVIDIA}},
  title   = {{NVIDIA Ethernet Networking Accelerates World's Largest AI Supercomputer, xAI Colossus}},
  year    = {2024},
  month   = oct,
  url     = {https://nvidianews.nvidia.com/news/spectrum-x-ethernet-networking-xai-colossus},
  urldate = {2026-02-17}
}

@online{oracle_supercluster_2024,
  author  = {{Oracle Cloud Infrastructure}},
  title   = {{Announcing World's Largest AI Supercomputer in the Cloud}},
  year    = {2024},
  month   = sep,
  url     = {https://blogs.oracle.com/cloud-infrastructure/worlds-largest-ai-supercomputer-in-the-cloud},
  urldate = {2026-02-17}
}

@article{liao2025ub,
  title={{UB-Mesh: A Hierarchically Localized nD-FullMesh Data Center Network Architecture}},
  author={Liao, Heng and Liu, Bingyang and Chen, Xianping and Guo, Zhigang and Cheng, Chuanning and Wang, Jianbing and Chen, Xiangyu and Dong, Peng and Meng, Rui and Liu, Wenjie and others},
  journal={IEEE Micro},
  year={2025},
  publisher={IEEE},
  volume = {45},
  number = {5},
  pages = {20--29},
  doi = {10.1109/mm.2025.3592688}
}

@misc{stanford2025aiindex,
  author       = {{Stanford Institute for Human-Centered Artificial Intelligence}},
  title        = {{The 2025 {AI} {Index} {Report}}},
  year         = {2025},
  howpublished = {\url{https://hai.stanford.edu/ai-index/2025-ai-index-report}},
  note         = {Accessed: 2026-05-31}
}

@misc{epoch2025trends,
  author       = {{Epoch AI}},
  title        = {{Data on {AI} Trends: Compute, Models, Hardware}},
  year         = {2025},
  howpublished = {\url{https://epoch.ai/trends}},
  note         = {Accessed: 2026-09-13}
}

@techreport{shehabi2025lbnl,
  author       = {Shehabi, Arman and Smith, Sarah J. and Hubbard, Alex and Newkirk, Alex
                  and Lei, Nuoa and Siddik, Md Abu Bakar and Holecek, Billie
                  and Koomey, Jonathan and Masanet, Eric and Sartor, Dale},
  title        = {{2024 {U.S.} Data Center Energy Usage Report}},
  institution  = {Lawrence Berkeley National Laboratory},
  year         = {2024},
  number       = {LBNL-2001637},
  howpublished = {\url{https://newscenter.lbl.gov/2025/01/15/berkeley-lab-report-evaluates-increase-in-electricity-demand-from-data-centers/}},
  note         = {Released January 2025}
}

@article{gholami2024memorywall,
  author    = {Gholami, Amir and Yao, Zhewei and Kim, Sehoon and Hooper, Coleman
               and Mahoney, Michael W. and Keutzer, Kurt},
  title     = {{{AI} and Memory Wall}},
  journal   = {IEEE Micro},
  volume    = {44},
  number    = {3},
  pages     = {33--39},
  year      = {2024},
  doi       = {10.1109/MM.2024.3373763},
  eprint    = {2403.14123},
  archivePrefix = {arXiv}
}

@misc{reuther2025laics,
  author       = {Reuther, Albert and Michaleas, Peter and Jones, Michael and
                  Gadepally, Vijay and Samsi, Siddharth and Kepner, Jeremy},
  title        = {{{LAICS}: {Lincoln} {AI} Computing Survey and Trends}},
  year         = {2025},
  eprint       = {2510.20931},
  archivePrefix = {arXiv},
  primaryClass = {cs.AR},
  howpublished = {\url{https://arxiv.org/abs/2510.20931}},
  note         = {IEEE HPEC 2025}
}

@misc{duan2024training,
  author       = {Duan, Jiangfei and Zhang, Shuo and Wang, Zerui and Jiang, Lijuan
                  and Qu, Wenwen and Hu, Qinghao and Wang, Guoteng and Weng, Qizhen
                  and Yan, Hang and Zhang, Xingcheng and Qiu, Xipeng and Lin, Dahua
                  and Wen, Yonggang and Jin, Xin and Zhang, Tianwei and Sun, Peng},
  title        = {{Efficient Training of Large Language Models on Distributed
                  Infrastructures: A Survey}},
  year         = {2024},
  eprint       = {2407.20018},
  archivePrefix = {arXiv},
  primaryClass = {cs.DC},
  howpublished = {\url{https://arxiv.org/abs/2407.20018}}
}

@misc{li2024llminfer,
  author       = {Li, Jinhao and Xu, Jiaming and Huang, Shan and Chen, Yonghua
                  and Li, Wen and Liu, Jun and Lian, Yaoxiu and Pan, Jiayi
                  and Ding, Li and Zhou, Hao and Wang, Yu and Dai, Guohao},
  title        = {{Large Language Model Inference Acceleration: A Comprehensive
                  Hardware Perspective}},
  year         = {2024},
  eprint       = {2410.04466},
  archivePrefix = {arXiv},
  primaryClass = {cs.AR},
  howpublished = {\url{https://arxiv.org/abs/2410.04466}}
}

@misc{xu2025nnsurvey,
  author       = {Xu, Bin and Banerjee, Ayan and Gupta, Sandeep K. S.},
  title        = {{Hardware Acceleration for Neural Networks: A Comprehensive Survey}},
  year         = {2025},
  eprint       = {2512.23914},
  archivePrefix = {arXiv},
  primaryClass = {cs.AR},
  howpublished = {\url{https://arxiv.org/abs/2512.23914}}
}

@misc{koilia2024llmhw,
  author       = {Koilia, Nikoletta and Kachris, Christoforos},
  title        = {{Hardware Acceleration of {LLMs}: A Comprehensive Survey and Comparison}},
  year         = {2024},
  eprint       = {2409.03384},
  archivePrefix = {arXiv},
  primaryClass = {cs.AR},
  howpublished = {\url{https://arxiv.org/abs/2409.03384}}
}

@article{silvano2025dlhpc,
  author    = {Silvano, Cristina and Ielmini, Daniele and Ferrandi, Fabrizio and Fiorin, Leandro and Curzel, Serena and Benini, Luca and Conti, Francesco and Garofalo, Angelo and Zambelli, Cristian and Calore, Enrico and Schifano, Sebastiano and Palesi, Maurizio and Ascia, Giuseppe and Patti, Davide and Petra, Nicola and De Caro, Davide and Lavagno, Luciano and Urso, Teodoro and Cardellini, Valeria and Cardarilli, Gian Carlo and Birke, Robert and Perri, Stefania},
  title     = {{A Survey on Deep Learning Hardware Accelerators for Heterogeneous
               {HPC} Platforms}},
  journal   = {ACM Computing Surveys},
  year      = {2025},
  doi       = {10.1145/3729215},
  volume = {57},
  number = {11},
  articleno = {286},
  numpages = {39},
  month = {jun}
}

@article{park2025engines,
  author    = {Park, Sihyeong and Jeon, Sungryeol and Lee, Chaelyn and
               Jeon, Seokhun and Kim, Byung-Soo and Lee, Jemin},
  title     = {{A Survey on Inference Engines for Large Language Models: Perspectives
               on Optimization and Efficiency}},
  journal   = {ACM Transactions on Intelligent Systems and Technology},
  year      = {2026},


  doi       = {10.1145/3803798},
  note = {Advance online publication, March 30, 2026}
}

@article{guo2024codesign,
  author       = {Guo, Cong and Cheng, Feng and Du, Zhixu and Kiessling, James
                  and Ku, Jonathan and Li, Shiyu and Li, Ziru and Ma, Mingyuan
                  and Molom-Ochir, Tergel and Morris, Benjamin and Shan, Haoxuan
                  and Sun, Jingwei and Wang, Yitu and Wei, Chiyue and Wu, Xueying
                  and Wu, Yuhao and Yang, Hao Frank and Zhang, Jingyang
                  and Zhang, Junyao and Zheng, Qilin and Zhou, Guanglei
                  and Li, Hai and Chen, Yiran},
  title        = {{A Survey: Collaborative Hardware and Software Design in the Era
                  of Large Language Models}},
  year         = {2025},


  journal = {IEEE Circuits and Systems Magazine},
  volume = {25},
  number = {1},
  pages = {35--57},
  doi = {10.1109/MCAS.2024.3476008},
  url = {https://doi.org/10.1109/MCAS.2024.3476008}
}

@article{weingram2023xccl,
  author    = {Weingram, Adam and Li, Yuke and Qi, Hao and Ng, Darren and
               Dai, Liuyao and Lu, Xiaoyi},
  title     = {{{xCCL}: A Survey of Industry-Led Collective Communication Libraries
               for Deep Learning}},
  journal   = {Journal of Computer Science and Technology},
  volume    = {38},
  number    = {1},
  pages     = {166--195},
  year      = {2023},
  doi       = {10.1007/s11390-023-2894-6}
}

@online{nvidiaRubin2026,
  author       = {{NVIDIA}},
  title        = {{{NVIDIA Vera Rubin NVL72} Specifications}},
  organization = {NVIDIA},
  year         = {2026},
  month        = jan,
  url          = {https://www.nvidia.com/en-us/data-center/vera-rubin-nvl72/},
  urldate      = {2026-09-20},
  note         = {Specification snapshot: 19.2 TB/s HBM and 3 TB/s NVLink per GPU}
}

@online{amdMI455X2026,
  author       = {{AMD}},
  title        = {{AMD Instinct MI455X GPUs}},
  organization = {Advanced Micro Devices, Inc.},
  year         = {2026},
  month        = jul,
  url          = {https://www.amd.com/en/products/accelerators/instinct/mi400/mi455x.html},
  urldate      = {2026-08-25}
}

@online{amdCDNA5,
  author       = {{AMD}},
  title        = {{AMD CDNA 5 Architecture}},
  organization = {Advanced Micro Devices, Inc.},
  year         = {2026},
  url          = {https://www.amd.com/en/technologies/cdna.html},
  urldate      = {2026-08-25}
}

@online{deepseekV4Pro,
  author       = {{DeepSeek-AI}},
  title        = {{DeepSeek V4 Preview Release}},
  organization = {DeepSeek},
  year         = {2026},
  month        = apr,
  url          = {https://api-docs.deepseek.com/news/news260424/},
  urldate      = {2026-08-25}
}

@online{kimiK3,
  author       = {{Moonshot AI}},
  title        = {{Kimi K3: Open Frontier Intelligence}},
  organization = {Moonshot AI},
  year         = {2026},
  month        = jul,
  url          = {https://www.kimi.com/en/blog/kimi-k3},
  urldate      = {2026-08-25}
}

@misc{nvidiaNCCLCollectives,
  author       = {{NVIDIA}},
  title        = {{Collective Operations}},
  organization = {{NVIDIA Collective Communication Library} Documentation},
  year         = {2026},
  url          = {https://docs.nvidia.com/deeplearning/nccl/user-guide/docs/usage/collectives.html},
  urldate      = {2026-08-26}
}

@misc{nvidiaNCCLAlgorithms,
  author       = {{NVIDIA}},
  title        = {{{NCCL} Collective Algorithm Selection}},
  organization = {{NVIDIA Collective Communication Library} Documentation},
  year         = {2026},
  url          = {https://docs.nvidia.com/deeplearning/nccl/user-guide/docs/env.html#nccl-algo},
  urldate      = {2026-08-26}
}

@inproceedings{rajbhandari2020zero,
  author    = {Rajbhandari, Samyam and Rasley, Jeff and Ruwase, Olatunji and He, Yuxiong},
  title     = {{{ZeRO}: Memory Optimizations Toward Training Trillion Parameter Models}},
  booktitle = {Proceedings of the International Conference for High Performance Computing, Networking, Storage and Analysis},
  series    = {{SC} '20},
  pages     = {1--16},
  year      = {2020},
  publisher = {IEEE Press},
  doi       = {10.1109/SC41405.2020.00024}
}

@misc{shoeybi2019megatron,
  author        = {Shoeybi, Mohammad and Patwary, Mostofa and Puri, Raul and LeGresley, Patrick and Casper, Jared and Catanzaro, Bryan},
  title         = {{{Megatron-LM}: Training Multi-Billion Parameter Language Models Using Model Parallelism}},


  year          = {2019},
  eprint        = {1909.08053},
  archivePrefix = {arXiv},
  url           = {https://arxiv.org/abs/1909.08053}
}

@article{korthikanti2023sequence,
  author  = {Korthikanti, Vijay Anand and Casper, Jared and Lym, Sangkug and McAfee, Lawrence and Andersch, Michael and Shoeybi, Mohammad and Catanzaro, Bryan},
  title   = {{Reducing Activation Recomputation in Large Transformer Models}},
  journal = {Proceedings of Machine Learning and Systems},
  volume  = {5},
  pages   = {341--353},
  year    = {2023},
  url     = {https://proceedings.mlsys.org/paper_files/paper/2023/hash/80083951326cf5b35e5100260d64ed81-Abstract-mlsys2023.html}
}

@inproceedings{lepikhin2021gshard,
  author    = {Lepikhin, Dmitry and Lee, HyoukJoong and Xu, Yuanzhong and Chen, Dehao and Firat, Orhan and Huang, Yanping and Krikun, Maxim and Shazeer, Noam and Chen, Zhifeng},
  title     = {{{GShard}: Scaling Giant Models with Conditional Computation and Automatic Sharding}},
  booktitle = {9th International Conference on Learning Representations ({ICLR})},
  year      = {2021},
  url       = {https://openreview.net/forum?id=qrwe7XHTmYb}
}

@article{patarasuk2009ring,
  author  = {Patarasuk, Pitch and Yuan, Xin},
  title   = {{Bandwidth Optimal All-Reduce Algorithms for Clusters of Workstations}},
  journal = {Journal of Parallel and Distributed Computing},
  volume  = {69},
  number  = {2},
  pages   = {117--124},
  year    = {2009},
  doi     = {10.1016/j.jpdc.2008.09.002}
}

@misc{jeaugey2019nccltrees,
  author       = {Jeaugey, Sylvain},
  title        = {{Massively Scale Your Deep Learning Training with {NCCL} 2.4}},
  organization = {{NVIDIA} Technical Blog},
  year         = {2019},
  month        = feb,
  url          = {https://developer.nvidia.com/blog/massively-scale-deep-learning-training-nccl-2-4/},
  urldate      = {2026-08-26}
}

@article{thakur2005mpich,
  author  = {Thakur, Rajeev and Rabenseifner, Rolf and Gropp, William},
  title   = {{Optimization of Collective Communication Operations in {MPICH}}},
  journal = {The International Journal of High Performance Computing Applications},
  volume  = {19},
  number  = {1},
  pages   = {49--66},
  year    = {2005},
  doi     = {10.1177/1094342005051521}
}

@misc{jeaugey2025pat,
  author        = {Jeaugey, Sylvain},
  title         = {{{PAT}: A New Algorithm for All-Gather and Reduce-Scatter Operations at Scale}},
  year          = {2025},
  eprint        = {2506.20252},
  archivePrefix = {arXiv},
  primaryClass  = {cs.DC},
  url           = {https://arxiv.org/abs/2506.20252}
}

@inproceedings{shah2023taccl,
  author    = {Shah, Aashaka and Chidambaram, Vijay and Cowan, Meghan and Maleki, Saeed and Musuvathi, Madan and Mytkowicz, Todd and Nelson, Jacob and Saarikivi, Olli and Singh, Rachee},
  title     = {{{TACCL}: Guiding Collective Algorithm Synthesis Using Communication Sketches}},
  booktitle = {20th USENIX Symposium on Networked Systems Design and Implementation ({NSDI} 23)},
  pages     = {593--612},
  year      = {2023},
  publisher = {{USENIX} Association},
  address   = {Boston, MA},
  isbn      = {978-1-939133-33-5},
  url       = {https://www.usenix.org/conference/nsdi23/presentation/shah}
}

@online{nvidiaCPO2025,
  author       = {Seyedi, Ashkan},
  title        = {{Scaling {AI} Factories with Co-Packaged Optics for Better
                  Power Efficiency}},
  organization = {{NVIDIA} Technical Blog},
  year         = {2025},
  month        = aug,
  url          = {https://developer.nvidia.com/blog/scaling-ai-factories-with-co-packaged-optics-for-better-power-efficiency/},
  urldate      = {2026-08-26}
}

@online{nvidia800VDC2025,
  author       = {Blake, Mathias and Hsu, Martin and Goldwasser, Ivan and
                  Petty, Harry and Huntington, Jared},
  title        = {{{NVIDIA} 800 {VDC} Architecture Will Power the Next
                  Generation of {AI} Factories}},
  organization = {{NVIDIA} Technical Blog},
  year         = {2025},
  month        = may,
  url          = {https://developer.nvidia.com/blog/?p=100571},
  urldate      = {2026-08-26}
}

@misc{choukse2025power,
  author        = {Choukse, Esha and Warrier, Brijesh and Heath, Scot and others},
  title         = {{Power Stabilization for {AI} Training Datacenters}},
  year          = {2025},
  eprint        = {2508.14318},
  archivePrefix = {arXiv},
  primaryClass  = {cs.AR},
  doi           = {10.48550/arXiv.2508.14318},
  url           = {https://arxiv.org/abs/2508.14318}
}

@online{openaiJalapeno2026,
  author       = {{OpenAI}},
  title        = {{{Jalape\~no}'s First Results Show Industry-Leading Speed and
                  Efficiency in {AI} Inference}},
  organization = {{OpenAI}},
  year         = {2026},
  month        = aug,
  url          = {https://openai.com/index/jalapeno-first-results/},
  urldate      = {2026-08-30}
}

@article{ma2026inferencehardware,
  author        = {Ma, Xiaoyu and Patterson, David},
  title         = {{Challenges and Research Directions for Large Language Model
                   Inference Hardware}},
  year          = {2026},


  doi           = {10.1109/MC.2026.3652916},
  url           = {https://doi.org/10.1109/MC.2026.3652916},
  journal = {Computer},
  volume = {59},
  number = {5},
  pages = {55--64}
}

@inproceedings{nair2026raptor,
  author    = {Nair, Prashant J. and Hadidi, Ramyad and Ganesh, Subramani and
               Kodge, Sangamesh and Rathore, Shubhankit and Thanawala, Neil and
               Reddy, Nikitha and Saharia, Gyanesh and Patankar, Vinayak and
               Tiruvur, Arun and Kurella, Nithesh and Bhoja, Sudeep},
  title     = {{Early Silicon of Raptor: The First 3D-{DRAM} Accelerator for
               Generative Inference}},
  booktitle = {2026 {ACM}/{IEEE} 53rd Annual International Symposium on
               Computer Architecture ({ISCA})},
  pages     = {2632--2647},
  year      = {2026},
  publisher = {{IEEE}},
  doi       = {10.1109/ISCA66397.2026.00183}
}

@online{nvidiaCoolingEfficiency2026,
  author       = {Moseley, Kibibi and Perez, Kristen and Mahajan, Pawini},
  title        = {{Scaling Token Factory Revenue and {AI} Efficiency by
                  Maximizing Performance per Watt}},
  organization = {{NVIDIA} Technical Blog},
  year         = {2026},
  month        = mar,
  url          = {https://developer.nvidia.com/blog/scaling-token-factory-revenue-and-ai-efficiency-by-maximizing-performance-per-watt/},
  urldate      = {2026-09-05}
}

@online{taalas2026ubiquitous,
  author       = {Bajic, Ljubisa},
  title        = {{The Path to Ubiquitous {AI}}},
  organization = {Taalas},
  year         = {2026},
  url          = {https://taalas.com/the-path-to-ubiquitous-ai/},
  urldate      = {2026-09-05}
}

@misc{chen2025griddemand,
  author        = {Chen, Xin and Wang, Xiaoyang and Colacelli, Ana and Lee, Matt and Xie, Le},
  title         = {{Electricity Demand and Grid Impacts of {AI} Data Centers: Challenges and Prospects}},
  year          = {2025},
  eprint        = {2509.07218},
  archivePrefix = {arXiv},
  primaryClass  = {eess.SY},
  doi           = {10.48550/arXiv.2509.07218}
}

@misc{bashir2026gridcodesign,
  author        = {Bashir, Noman and Sherwood, Rob and Xie, Le and Yu, Minlan},
  title         = {{From Barrier to Bridge: The Case for {AI} Data Center/Power Grid Co-Design}},
  year          = {2026},
  eprint        = {2605.03090},
  archivePrefix = {arXiv},
  doi           = {10.48550/arXiv.2605.03090}
}

@online{googleTPUComparison2026,
  author       = {{Google Cloud}},
  title        = {{{TPU} Machines in Accelerator-Optimized Machine Family}},
  year         = {2026},
  url          = {https://docs.cloud.google.com/compute/docs/tpus/tpu-machines},
  urldate      = {2026-09-07}
}

@online{nvidiaNVL72Components,
  author       = {{NVIDIA}},
  title        = {{System Hardware \& Components: {NVIDIA NVL72 AI Factory}}},
  year         = {2026},
  url          = {https://docs.nvidia.com/enterprise-reference-architectures/nvl72-ai-factory/latest/components.html},
  urldate      = {2026-09-07}
}

@online{googleTPUIntroduction2016,
  author       = {{Google}},
  title        = {{Google Supercharges Machine Learning Tasks with Custom Chip}},
  year         = {2016},
  month        = may,
  url          = {https://cloud.google.com/blog/products/ai-machine-learning/google-supercharges-machine-learning-tasks-with-custom-chip},
  urldate      = {2026-09-12}
}

@online{awsInferentiaIntroduction2018,
  author       = {{Amazon Web Services}},
  title        = {{Announcing {Amazon Inferentia}: A Machine Learning Inference Microchip}},
  year         = {2018},
  month        = nov,
  url          = {https://aws.amazon.com/about-aws/whats-new/2018/11/announcing-amazon-inferentia-machine-learning-inference-microchip/},
  urldate      = {2026-09-12}
}

@online{microsoftMaiaIntroduction2023,
  author       = {{Microsoft}},
  title        = {{Microsoft Ignite 2023: {AI} Transformation and the Technology Driving Change}},
  year         = {2023},
  month        = nov,
  url          = {https://blogs.microsoft.com/blog/2023/11/15/microsoft-ignite-2023-ai-transformation-and-the-technology-driving-change/},
  urldate      = {2026-09-12}
}

@online{awsTrainiumIntroduction2020,
  author       = {Barr, Jeff},
  title        = {{re:Invent 2020 Liveblog: Andy Jassy Keynote}},
  year         = {2020},
  month        = dec,
  url          = {https://aws.amazon.com/blogs/aws/reinvent-2020-liveblog-andy-jassy-keynote/},
  urldate      = {2026-09-12}
}

@inproceedings{metaMTIA300ISCA2026,
  author       = {{MTIA Team}},
  title        = {{{MTIA 300}: Meta's First Training Chip Featuring Built-in {NICs} and Collective Offloading Engines}},
  booktitle    = {Proceedings of the 53rd Annual International Symposium on Computer Architecture},
  year         = {2026},
  url          = {https://aisystemcodesign.github.io/papers/MTIA300_ISCA2026.pdf},
  urldate      = {2026-09-13},
  pages = {1084--1099},
  doi = {10.1109/isca66397.2026.00085},
  publisher = {IEEE},
  note = {Group authorship follows the author-distributed paper}
}

@online{metaMTIAEvolution2026,
  author       = {Song, Yee Jiun and Tulloch, Andrew and Reddy, Harikrishna and Tang, CQ and Thakkar, Vijay},
  title        = {{Four {MTIA} Chips in Two Years: Scaling {AI} Experiences for Billions}},
  year         = {2026},
  month        = mar,
  url          = {https://ai.meta.com/blog/meta-mtia-scale-ai-chips-for-billions/},
  urldate      = {2026-09-12}
}

@online{microsoftMaia200Introduction2026,
  author       = {Guthrie, Scott},
  title        = {{{Maia 200}: The {AI} Accelerator Built for Inference}},
  year         = {2026},
  month        = jan,
  url          = {https://blogs.microsoft.com/blog/2026/01/26/maia-200-the-ai-accelerator-built-for-inference/},
  urldate      = {2026-09-12}
}

@online{graphcorePod64,
  author = {{Graphcore}},
  title = {{{IPU-POD64} Reference Design Datasheet}},
  year = {2020},
  note = {Version 1.0, December 2, 2020; IPU-Link topology},
  url = {https://docs.graphcore.ai/projects/ipu-pod64-datasheet/en/1.0.0/_static/GC-000477-DS-6-IPU-POD64-datasheet.pdf},
  urldate = {2026-09-20}
}

@online{amdMI100Brief,
  author = {{AMD}},
  title = {{{AMD Instinct MI100} Accelerator}},
  year = {2020},
  url = {https://www.amd.com/content/dam/amd/en/documents/instinct-business-docs/product-briefs/instinct-mi100-brochure.pdf},
  urldate = {2026-09-20}
}

@online{amdMI355XDatasheet,
  author = {{AMD}},
  title = {{{AMD Instinct MI355X GPU}: Leading-Edge {GPU} for Generative {AI}, Inference, Training, and High Performance Computing}},
  year = {2025},
  note = {GPU datasheet, available from the product documentation; dense FP16/BF16 2.5166 PFLOPS, FP8/INT8 5.0332 Peta-operations/s},
  url = {https://www.amd.com/en/products/accelerators/instinct/mi350/mi355x.html},
  urldate = {2026-09-20}
}

@online{amdMI50Specs,
  author = {{AMD}},
  title = {{Accelerator Specifications: {Radeon Instinct MI50} (32 {GB})}},
  note = {Undated product specifications; 184 GB/s total Infinity Fabric bandwidth},
  url = {https://www.amd.com/en/products/specifications/accelerators.html},
  urldate = {2026-09-20}
}

@online{cambriconMLU270,
  author = {{Cambricon}},
  title = {{{MLU270} Product Specifications}},
  year = {2019},
  url = {https://www.cambricon.com/index.php?m=content&c=index&a=lists&catid=37},
  urldate = {2026-09-20}
}

\appendix
\section{AI-Assisted Language Polishing}
\label{app:language-polishing}

OpenAI Codex was used to assist with language polishing, including spelling, grammatical agreement, sentence structure, and terminology consistency. 

\end{document}